\documentclass[%
 reprint,
 amsmath,amssymb,
 aps,
 english
]{revtex4-1}

\usepackage{setspace}
\usepackage[utf8]{inputenc} 
\usepackage[english]{babel} 
\usepackage{graphicx}
\usepackage{latexsym}
\usepackage{color}
\usepackage[table]{xcolor}
\usepackage{amsmath,amsthm,amssymb}
\usepackage{accents}
\usepackage{flafter}
\usepackage{amssymb}
\usepackage{bbold} 
\usepackage{bm}
\usepackage{pstricks}
\usepackage{pst-plot}
\usepackage{cancel}
\usepackage{eufrak,mathrsfs,mathrsfs}
\usepackage{dsfont}    
\usepackage{makeidx}         
\usepackage{nomencl}         
\usepackage{float}
\usepackage{xfrac}
\usepackage{hyperref}
\usepackage{xcolor}

\definecolor{red}{rgb}{1.0,0.0,0.0}
\definecolor{blue}{rgb}{0.0,0.0,1.0}
\definecolor{dark-gree}{rgb}{0.0,0.5,0.0}
\hypersetup{
    colorlinks, 
    linkcolor={red},
    citecolor={blue}, 
    urlcolor={dark-gree}
}

\usepackage{graphicx}
\usepackage{dcolumn}
\usepackage{bm}

\newcommand{\eq}[1]{\begin{equation}#1\end{equation}}
\newcommand{\tn}[1]{\textnormal{#1}}
\newcommand{\lb}[1]{\label{#1}}
\newcommand{\dt}[1]{\accentset{\bullet}{#1}}

\newcommand{\bra}{\langle}
\newcommand{\ket}{\rangle}

\newcommand\norm[1]{\left\lVert#1\right\rVert}

\makeatletter
\newsavebox\myboxA
\newsavebox\myboxB
\newlength\mylenA
\newcommand*\xoverline[2][0.75]{%
    \sbox{\myboxA}{$\m@th#2$}%
    \setbox\myboxB\null
    \ht\myboxB=\ht\myboxA%
    \dp\myboxB=\dp\myboxA%
    \wd\myboxB=#1\wd\myboxA
    \sbox\myboxB{$\m@th\overline{\copy\myboxB}$}
    \setlength\mylenA{\the\wd\myboxA}
    \addtolength\mylenA{-\the\wd\myboxB}%
    \ifdim\wd\myboxB<\wd\myboxA%
       \rlap{\hskip 0.5\mylenA\usebox\myboxB}{\usebox\myboxA}%
    \else
        \hskip -0.5\mylenA\rlap{\usebox\myboxA}{\hskip 0.5\mylenA\usebox\myboxB}%
    \fi}

\begin{document}

\preprint{APS/123-QED}

\title{Estimating the time evolution of NMR systems via quantum speed limit-like expression}

\author{D. V. Villamizar}
 \email{david.velasco.v@gmail.com}
\author{E. I. Duzzioni}
 \email{duzzioni@gmail.com}
\affiliation{ Departamento de F\'isica, Universidade Federal de Santa Catarina, Santa Catarina, CEP 88040-900, Brazil}%
\author{A. C. S. Leal}
\author{R. Auccaise}
\affiliation{ Departamento de F\'isica, Universidade Estadual de Ponta Grossa, Av. Carlos Cavalcanti, 4748, Ponta Grossa, Paran\'{a}, CEP 84030-900, Brazil}%

\date{\today}

\begin{abstract}
Finding the solutions of the equations that describe the dynamics of a given physical system is crucial in order to obtain important information about its evolution. However, by using estimation theory, it is possible to obtain, under certain limitations, some information on its dynamics. The quantum-speed-limit (QSL) theory was originally used to estimate the shortest time in which a Hamiltonian drives an initial state to a final one for a given fidelity. Using the QSL theory in a slightly different way, we are able to estimate the running time of a given quantum process. For that purpose, we impose the saturation of the Anandan-Aharonov bound in a rotating frame of reference where the state of the system travels slower than in the original frame (laboratory frame). Through this procedure it is possible to estimate the actual evolution time in the laboratory frame of reference with good accuracy when compared to previous methods. Our method is tested successfully to predict the time spent in the evolution of nuclear spins 1/2 and 3/2 in NMR systems. We find that the estimated time according to our method is better than previous approaches by up to four orders of magnitude. One disadvantage of our method is that we need to solve a number of transcendental equations, which increases with the system dimension and parameter discretization used to solve such equations numerically.
\end{abstract}

\maketitle
\section{Introduction}
We can address the problem we are tackling by means of asking ourselves a question: Can we know the time interval of a quantum process without solving its relevant dynamical equations? An affirmative answer to this question would be of great importance in situations in which is hard to solve both the Schr\"odinger (nonrelativistic) and Dirac (relativistic) equations. Yet, in general, the solution for Hamiltonians (i.e., equations that describe the time evolution of a system) that show an explicit time dependence or that take into account many-body interactions is even harder to solve. In this work we address this problem in the particular case in which the pure initial quantum state evolves unitarily in time. Our approach is based on a variation of the quantum-speed-limit (QSL) time, i.e., the minimum time required for a quantum system to evolve from an initial state to a final one.

In the context of the energy-time uncertainty relations, Mandelstam and Tamm (MT) \cite{Mandelstam_1945} reported a QSL time $t \!\geq\! \hbar  \arccos \sqrt{F(t)}/\Delta H$ for a closed quantum system evolving between two distinct pure states $|\psi (0)\ket$ and $|\psi(t)\ket$, where $t$ is the actual time of the evolution. Such expression is valid for a time-independent Hamiltonian $\hat{H}$ with energy uncertainty given by $\Delta H \!=\! \sqrt{\bra \psi (t) |\hat{H}^2 | \psi(t)\ket - \bra \psi(t)|\hat{H}| \psi(t) \ket^2}$ and the fidelity between the initial and final states defined by
\eq{ \label{eq:fidelity}
   F(t) = \big| \bra \psi (0)|\psi(t)\ket \big|^2.
}
Later, Margolus and Levitin \cite{Margolus_1998} resorting to energy as a resource, developed an alternative expression for the QSL time $t \!\geq\! h/4( \bra \hat{H} \ket \!-\! E_0 ) $, where the term in parentheses in the denominator means the average energy above the energy of the reference state. Since these two remarkable works, intense study on this subject emerged, including generalizations for unitary \cite{Fleming_1973,Bhattacharyya_1983,Anandan_1990,Vaidman_1992,Uhlmann_1992,Uffink_1993,Pfeifer_1993,Pfeifer_1995,Giovannetti_2003,Levitin_2009,Deffner_2013a,Marvian_2016} and nonunitary \cite{Taddei_2013,Campo_2013,Deffner_2013b,Zhang_2014,Pires_2016} evolutions of quantum states. 

\begin{figure}[h!]
 \centering
 \includegraphics[scale=0.055]{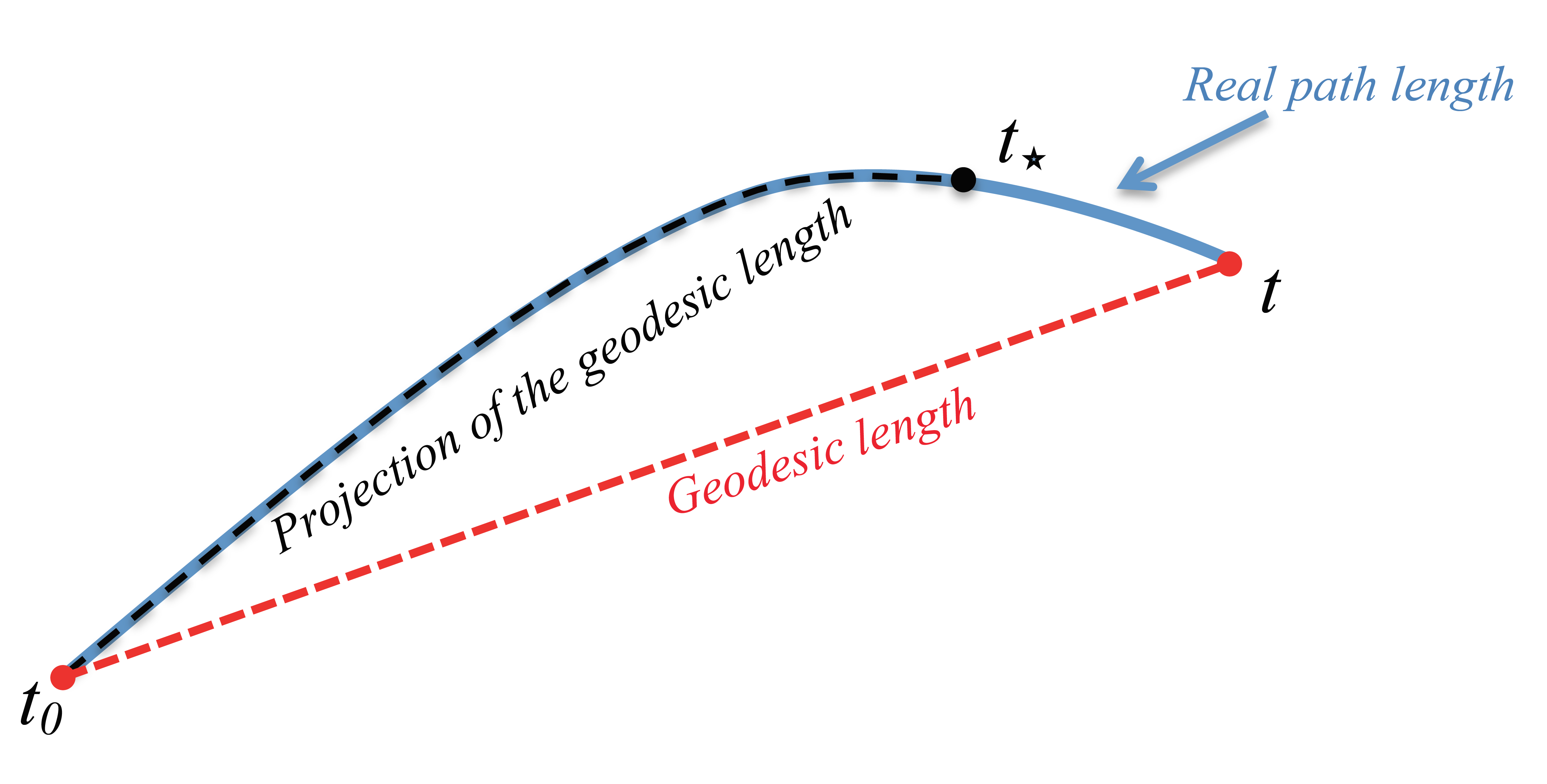}
 \caption{\small (Color online) Illustration of the geometric meaning of the AAT $t_{\star}$ defined according to Eq. (\ref{eq:qsl}). $t_{\star}$ is the time for the system state traveling at 
average speed $ \xoverline{\Delta H(t)}/\hbar$ to cross over the geodesic length $\arccos \sqrt{F (t)}$ between the initial and final states with fidelity given by Eq. (\ref{eq:fidelity}). $t$ is the total evolution time.}
 \label{fig:qsl}
\end{figure}

On this subject we woud like to highlight the seminal work of Anandan and Aharonov \cite{Anandan_1990}, which proposed a geometrical approach to the state evolution in quantum mechanics. Using the Fubini-Study metric in the projective Hilbert space, they found the shortest path between distinct pure states and also determined the average speed of the state evolution through the energy uncertainty. The ratio between these two quantities gave origin to the following expression
\begin{equation} \label{eq:qsl}
   t \geqslant t_{\star} = \frac{\hbar \arccos \sqrt{F(t)}}{\xoverline{\Delta H(t)}},
\end{equation}
 where the term $\arccos \! \sqrt{F(t)}$ is the geodesic distance between the initial $|\psi (0) \ket$ and final $|\psi (t)\ket$ states. As the Hamiltonian can be time dependent now, we are able to define the average speed of the state evolution $\frac{\small{\xoverline{\Delta H(t)}}}{\hbar} \equiv \frac{1}{t}\int_0^t \frac{\Delta H(\tau)}{\hbar} d\tau$ \cite{Anandan_1990}. Due to its beautiful geometrical interpretation \cite{Anandan_1990,Uhlmann_1992,Taddei_2013,Pires_2016}, as shown in Figure \ref{fig:qsl}, the expression for the average time of the state evolution (\ref{eq:qsl}) will be used henceforth and $t_{\star}$ will be called Anandan-Aharonov time (AAT).

Although the expression above has been interpreted as an estimation of the shortest time of a given quantum evolution, i.e., the QSL time, in Ref. \cite{Mirkin_2016} it was shown for nonunitary dynamics that expressions that depend on the average speed of the state evolution are in fact estimating the actual time of the evolution instead of the shortest one. The formulas developed in Ref. \cite{Deffner_2013b} are good examples of such case. 

Here we apply Eq. (\ref{eq:qsl}) to estimate the actual time for an initial pure quantum state in the laboratory frame to achieve a certain fidelity under the action of a unitary evolution generated by a time-dependent Hamiltonian. For this purpose, we recall that the AAT becomes the actual time when the quantum state evolves over the geodesic path \cite{Taddei_2013,Pires_2016}. The main idea of this work is to move the description to a different reference frame where the quantum state performs a path which is closer to the geodesic path, therefore evolving slower than in the original reference frame. We can use this fact to impose the saturation of the Anandan-Aharonov (AA) bound (\ref{eq:qsl}) in a rotating frame in which the Hamiltonian is time independent. Such an imposition transforms Eq. (\ref{eq:qsl}) into a transcendental equation, where its solutions enable us to estimate the actual time at which the initial state of the quantum system in the laboratory frame evolves until achieving a certain fidelity $F$. As we moved the system to a frame where the Hamiltonian is time independent, the average speed of the evolution became constant. This means that in the rotating frame Eq. (\ref{eq:qsl}) becomes useful, since the average speed of the state evolution does not depend on the unknown actual evolution time \cite{Mirkin_2016}.

We organize this paper as follows. Section \ref{sec:method} briefly explains the action of a unitary transformation over the time-dependent Schr\"odinger equation to obtain a frame of reference in which we describe the evolution of the quantum state and calculate its energy uncertainty. Also, we show a way to express the fidelity of the evolved state (unknown) of the system in this frame. Through the use of the AAT, we propose a method to estimate the dynamics of the quantum system in the laboratory frame according to its initial state. In Sec. \ref{sec:applications} we apply the proposed method in two NMR systems of spins 1/2 and 3/2, respectively. We compare the time evolution predicted by our method with the theoretical predictions of the actual evolution time made by the integration of the Schr\"odinger equation and the experimental data. In Sec. \ref{sec:discussions} we discuss the results and present our conclusions.
\section{The method} \label{sec:method}
Let us consider that in the laboratory frame of reference a quantum system is described by the time-dependent Schr\"odinger equation $i\hbar |\dt{\psi}(t) \ket \!=\! \hat{H}(t)|\psi(t) \ket$, where $\hat{H}(t)$ and $|\psi(t)\ket$ are the Hamiltonian and the quantum state of the system at time $t$, respectively. Given an initial state $|\psi(0)\ket$ and the system Hamiltonian, our main goal is to estimate the actual evolution time of the system to achieve a final state with fidelity $F(t)$. Notice that we do not know the final state of the evolution $|\psi(t)\ket$, just the value of the fidelity, which is given \emph{a priori}. From Eq. (\ref{eq:qsl}) we observe that $t \!=\! t_{\star}$ only when the evolution of the state of the system occurs over the geodesic path. In such a case, the quantum state progresses as slowly as possible for a fixed time interval, which is equivalent to attaining the smallest value for the average energy uncertainty $ \xoverline{\Delta H(t)}$. Although the AAT values are different when evaluated in different frames of reference, a unitary transformation from the original frame to any other yields the same value for the actual evolution time. This can be seen through the application of a time-dependent unitary transformation $R(t)$, such that the quantum state can be expressed as $|\psi(t)\ket \!=\! R(t)|\psi_R(t)\ket$, satisfying the initial condition $|\psi(0) \ket \!=\! |\psi_R(0)\ket$. The quantum state $|\psi_R(t)\ket$ is represented in the new frame of reference at time $t$ and it evolves according to the time-dependent Schr\"odinger equation $i\hbar |\dt{\psi}_R(t) \ket = \hat{H}_R(t)|\psi_R(t) \ket$, where the Hamiltonian in the new frame is defined as $\hat{H}_R(t) \!=\! R^{\dagger}(t)\hat{H}(t)R(t) \!-\! i\hbar R^{\dagger}(t)\dt{R}(t)$. In particular, to estimate the actual time t, we will use the expression (\ref{eq:qsl}) in a specific rotating frame 
\begin{equation}\label{eq:qslRotating}
   t \geqslant t_{\star}^R = \frac{\hbar \arccos \sqrt{F_R(t)}}{\xoverline{\Delta H_R}},
\end{equation}
where the average energy uncertainty $\xoverline{\Delta H_R}$ is chosen to be smaller than its counterpart in the laboratory frame $\xoverline{\Delta H(t)}$. Hence, we hope to find $t^R_{\star}$ closer to the actual time $t$ when compared to the same prediction made in the laboratory frame, [see Eq. (\ref{eq:qsl})]. Choosing an appropriate unitary transformation $R(t)$, the Hamiltonian in the rotating frame $\hat{H}_R$ becomes time independent and consequently its average variance, $\xoverline{\Delta H_R(t)}=\xoverline{\Delta H_R}$. The later imposition was made for the sake of simplicity and also because the average energy uncertainty becomes time independent, i.e., it does not depend on the actual time of the evolution. Naturally, the expression for the fidelity in the rotating frame also changes, $F_R(t) \!=\! | \bra \psi_R(0)|\psi_R (t) \ket |^2 \!=\! |\bra \psi(0)|R^{\dagger}(t)|\psi (t) \ket |^2$.\\
%
\subsection{Motivation}
To understand the idea behind our method, we rewrite Eq. (\ref{eq:qsl}) as
\begin{equation} \label{eq:AAT_geom_1} 
t_{\star}=\frac{\ell_{geodes}}{\overline{v}},
\end{equation}
where $\ell_{\tn{geodes}}$ is the geodesic path length and $\overline{v}$ is the average speed of the quantum state evolution, which is taken over the whole evolution time. On the other hand, the average speed can be written as
\begin{equation} \label{eq:AAT_geom_2}
\overline{v}=\frac{\ell_{real}}{t},
\end{equation}
with $\ell_{\tn{real}}$ being the real path length and $t$ the actual evolution time. Therefore, the AAT is expressed as
\begin{equation} \label{times}
t_{\star}=\frac{\ell_{\tn{geodes}}}{\ell_{\tn{real}}}t.
\end{equation}
Once the actual time is the same in all frames of reference, we have only to find a reference frame where we saturate as much as possible the ratio $\ell_{\tn{geodes}}/\ell_{\tn{real}} \to 1$  in order to obtain a good estimate of the actual evolution time. According to our proposal, this is achieved  when the quantum state evolves as slower as possible in a rotated frame of reference, once Eqs. (\ref{eq:AAT_geom_1}) and (\ref{eq:AAT_geom_2}) remain valid for all frames of reference. We observe such effect by noticing in Eq. (\ref{eq:AAT_geom_2}) that small average speeds $\overline{v}$ implies small lengths, so that such lengths cannot be smaller than the geodesic one. This is illustrated in a video in the Supplemental Material \cite{SuppMat}. For the reasons stated above, we are led to choose a reference frame where the system Hamiltonian is time independent. This choice can be better understood by referring to the definition of the average speed of the state evolution and the fact that the energy uncertainty is time independent. The following are consequences of this choice. (i) We cannot guarantee that this is the optimal frame of reference where the speed of the evolution is as slow as possible and therefore that the system state is running over the geodesic length. However, for driven quantum systems such as NMR, cavity QED, cold atoms, ion traps, and superconducting quantum information devices \cite{Chen_2007}, where the system Hamiltonian in the Schrödinger picture is highly oscillating, our method may become useful, since in this frame of reference the energy uncertainty is smaller than in the laboratory frame, which is enough to improve the estimation of the actual evolution time. See the next paragraph to follow the demonstration of this last statement. (ii) There are situations in which is impossible to find a frame of reference where the Hamiltonian is time independent. When the transformed Hamiltonian $H_R(t)$ is necessarily time dependent and consequently its variance, the average speed of the state evolution in this frame of reference depends on the total evolution time, i.e., the variable to be estimated. An alternative to solve this problem would be performing an estimate of $\xoverline{\Delta H_R(t)}/\hbar$. The latter case will not be dealt with here and is left for future research. \\
    
Let us introduce a proper example to explain why moving to a frame of reference where the state evolution is slower than in the laboratory frame is suitable to estimate the actual evolution time. Consider a time-independent Hamiltonian $\hat{H}$ in the laboratory frame. We can choose the unitary transformation connecting to the rotating frame as
\begin{equation} \label{Eq_TUR}
R(t)=e^{-i\hat{H}\varepsilon^{\prime} t/\hbar}, \hspace{1cm} (\varepsilon^{\prime} \in \mathbb{R}), 
\end{equation}
so that in the new frame the Hamiltonian is $\hat{H}_{R}=\varepsilon \hat{H}$, where $\varepsilon \equiv  1-\varepsilon^{\prime}$. With the evolved state given by $|\psi_{R}(t)\ket \!=\! e^{-i\hat{H}\varepsilon t/\hbar}|\psi(0)\ket$, we obtain the fidelity
\begin{equation} \lb{eq_fidR}
F_R(t)=|\bra \psi(0) |e^{-i\hat{H}\varepsilon t/\hbar}| \psi(0)\ket|^2
\end{equation}
and the energy uncertainty
\begin{equation}\lb{eq_HamiltonianRTI}
\Delta H_R=\varepsilon\Delta H_0,
\end{equation}
with $\Delta H_0=\sqrt{\bra \psi (0) |\hat{H}^2 | \psi(0)\ket - \bra \psi(0)|\hat{H}| \psi(0) \ket^2}$ the energy uncertainty in the laboratory frame. From Eqs. (\ref{eq:qslRotating}), (\ref{eq_fidR}), and (\ref{eq_HamiltonianRTI}), we find the following expression for $t^R_{\star}$
\begin{equation} \lb{eq_tstarR}
t_{\star}^R = \frac{\hbar \arccos\left( |\bra \psi(0) |e^{-i\hat{H}\varepsilon t/\hbar}| \psi(0)\ket|  \right)}{\varepsilon\Delta H_0}.
\end{equation}
As we can find a frame of reference in which $\varepsilon$ can be arbitrary small, the fidelity $F_R(t)$ can be expanded in power series of this parameter as
\[
F_R(t) = 1 - \left( \frac{\varepsilon t}{\hbar} \right)^2\Delta H_0^2 + \mathcal{O}(\varepsilon^4).
\]
By performeing a similar procedure to write the function $\arccos$, Eq. (\ref{eq_tstarR}) becomes
\begin{align} 
t_{\star}^R &= \frac{\hbar}{\varepsilon\Delta H_0} \left( \frac{\varepsilon t \Delta H_0}{\hbar} + \mathcal{O}(\varepsilon^3)  \right) \nonumber \\
                 &=t+\mathcal{O}(\varepsilon^2).
\end{align}
We observe in Eq. (\ref{eq_HamiltonianRTI}) that the speed of the state evolution in the rotating frame is controlled by the parameter $\varepsilon$. As smaller values of such parameter imply slower state evolutions, then we obtain $\lim_{\varepsilon \to 0} t_{\star}^R = t$.\\
%
\subsection{Implementation}
To implement our method, first we notice that even for time-independent Hamiltonians in the rotating frame, the utility of Eq. (\ref{eq:qslRotating}) remains limited, provided that it is necessary to know the quantum state $|\psi(t)\ket$ to obtain the fidelity $F_R (t)$. In order to overcome this problem, we replace $F_R (t)$ by $\xoverline{F}_R(t^R_{\star})$ in Eq. (\ref{eq:qslRotating}), resulting in the transcendental equation 
\begin{equation}\label{eq:transcendental}
  \arccos \sqrt{\xoverline{F}_R(t^R_{\star})} = \frac{\Delta H_R \; t^R_{\star}}{\hbar}.
\end{equation}
The expression for the fidelity in the rotating frame $\xoverline{F_R}(t^R_{\star})$ is
\begin{equation}\lb{eq:fidelity_rotated}
  \xoverline{F_R}(t^R_{\star}) \equiv | \bra \psi(0)|R^{\dagger}(t^R_{\star})| \xoverline{\psi}(t^R_{\star}) \ket |^{2}, 
\end{equation}
with the quantum state at time $t^R_{\star}$ being written as
\begin{equation} \label{eq:state_expansion}
|\xoverline{\psi} (t^R_{\star}) \ket = \sqrt{F}|\psi(0)\ket + \sum^{n-1}_{j=1}a_{j}e^{i\varphi_{j}}|\psi^{\perp}_{j}\ket.
\end{equation}
This fidelity is a mathematical artifice to replace the actual evolution time $t$ by its estimate $t^R_{\star}$. A qualitative error analysis introduced by the replacement of $F_R(t)$ by $\xoverline{F}_R(t^R_{\star})$ is made in Appendix \ref{app:error_fidelity}. The quantum states $|\psi^{\perp}_{j}\ket$ are orthogonal to $|\psi(0)\ket$, the time-dependent coefficients $a_j,\varphi_j \!\in\! \mathbb{R}$ satisfy $\sum^{n-1}_{j=1}|a_{j}|^2 \!=\! 1\!-\!F$, and $n$ is the dimension of the Hilbert space. The reason we have written the evolved quantum state $|\psi(t)\ket$ as in Eq. (\ref{eq:state_expansion}) is that it recovers the expression (\ref{eq:fidelity}), i.e., $F \!=\! | \bra \psi(0)|\xoverline{\psi} (t^R_{\star}) \ket |^2$. Therefore, we can express $\xoverline{F}_R(t^R_{\star})$ as
\eq{\lb{eq:estimativa_FRG}
 \begin{split}
\xoverline{F}_R(t^R_{\star}) &= \Bigg|\sqrt{F} \bra\psi(0)|R^{\dagger}(t^R_{\star})|\psi(0)\ket \\
	&\hspace{1.0cm}+ \sum^{n-1}_{j=1}a_{j}e^{i\varphi_{j}} \bra\psi(0)|R^{\dagger}(t^R_{\star})|\psi^{\perp}_{j}\ket \Bigg|^{2}.
 \end{split}
}
We can obtain the set of $n\!-\!1$ orthogonal states by using the Gram-Schmidt orthogonalization procedure over the initial state. Hence, given the initial quantum state, the fidelity $F$, and the Hamiltonian in the laboratory frame, and after we have chosen the time-dependent unitary transformation $R(t)$, we find numerically the root of Eq. (\ref{eq:transcendental}) for the shortest time $t^R_{\star}$. We observe that the transcendental equation depends on several variables $\{\varphi_1,\dots, \varphi_{n-1},a_1,\dots,a_{n-1}\}$, beyond $t^R_{\star}$. Our strategy to solve such an equation is to sweep over the whole set of parameters to minimize $t^R_{\star}$ as the first chronological root of the transcendental equation. We are looking for the first chronological root because different final states can have the same value of fidelity. This may occur mainly in systems with periodic dynamics. It is also possible to obtain numerically the best unitary transformation that leads to the best time estimation, although this is not our goal here. This can be reached numerically by sweeping over all the parameters of the unitary transformation.

Although in this work we apply our method just for time-dependent Hamiltonians in NMR systems, it works as well as for time-independent Hamiltonians. In the latter case the unitary transformation must be of the form (\ref{Eq_TUR}) and the rest of the procedure remains.


Nonetheless, to analyze how effective the method proposed here is, in the next section we show its application to estimate the necessary time for an initial state to evolve to a final state in systems consisting of spins 1/2 and 3/2 in the NMR scenario. Our predictions are compared to experimental data for both evolutions. 
\section{Applications}\lb{sec:applications}
\subsection{Spin 1/2 systems}
In order to put in practice our method, we implemented experimentally the dynamics of a spin $I\!=\!1/2$ NMR system composed by molecules of o-phosphoric acid (see Appendix \ref{sec:level1}). Phosphorous nuclei ($^{31}$P) interact with an external magnetic field according to the Hamiltonian,
\eq{ \lb{eq:hamiltonian_1/2}
\hat{H}(t)=\hbar\big\{ \omega_0\hat{\mathbf{I}}_z +\omega_1\big[ \cos(\omega_p t) \hat{\mathbf{I}}_x + \sin(\omega_p t)\hat{\mathbf{I}}_y \big] \big\},
}
where $\omega_0$ is the Larmor frequency and $\omega_1$ is proportional to the intensity of the magnetic field applied in the $x\!-\!y$ plane which rotates around the $z$ axis with frequency $\omega_p$, while $\hat{\mathbf{I}}_i \!=\! \hat{\sigma}_i/2$, $i \!=\! x, y, z$, and $\hat{\sigma}_i$ are the Pauli spin matrices. We observe that the Hamiltonian (\ref{eq:hamiltonian_1/2}) does not commute with itself for different times, which introduces some degree of difficulty to solve the Schr\"odinger equation, despite its simple form.

\begin{figure}[t!]
 \centering
  \includegraphics[scale=0.38]{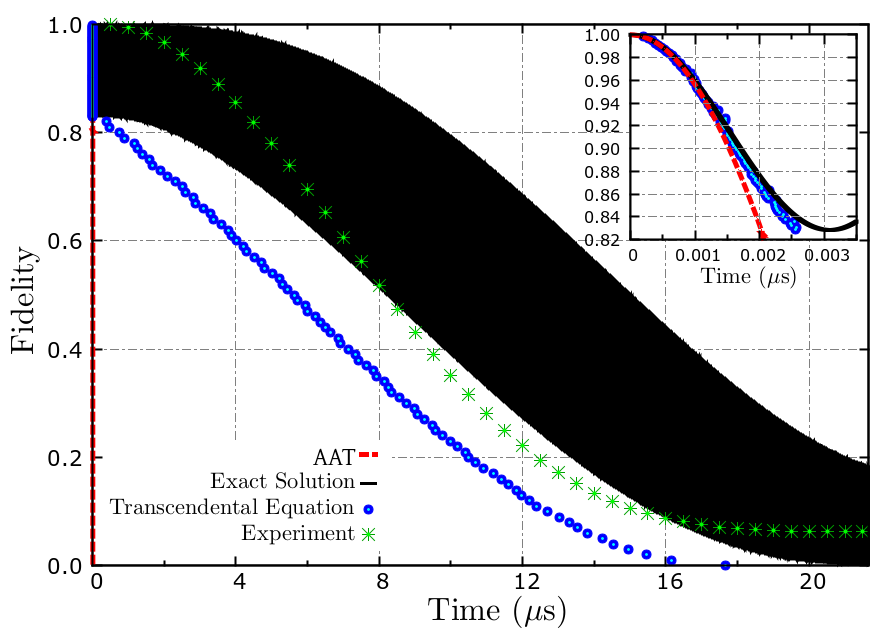}
 \caption{\small (Color online) The time need for the nuclear spin 1/2 of the $^{31}$P atom starting from the initial Bloch state characterized by the angles $\theta \!=\! 24.48^{\circ}$ and $\phi \!=\! 4.02^{\circ}$ to achieve a final state with time-dependent fidelity. The experimental measurements are the green asterisks, while the theoretical description is given by the solid black line, which seems a black belt due to its very fast oscillating behavior. See the inset to observe the time scale of the oscillation. The predictions of the actual evolution time for the system state to attain a desired fidelity are described by the AAT (red dashed line) and the first chronological roots of the transcendental equation (blue circles).}
 \label{fig:qsl_1/2}
\end{figure}

The unitary transformation that removes the time dependence from the Hamiltonian (\ref{eq:hamiltonian_1/2}) is $R(t) \!=\! e^{-i \hat{\mathbf{I}}_{z}\omega_{p}t}$, which is a rotation around the $z$ axis, explaining why we called it rotating frame. The resulting Hamiltonian is $\hat{H}_R \!=\! \hbar \big( \Delta \hat{\mathbf{I}}_z + \omega_1 \hat{\mathbf{I}}_x \big)$, with  $\Delta\!=\! \omega_0 \!-\! \omega_p$. We considered as the initial state of the system a general pure state on the Bloch sphere $|\psi(0)\ket \!=\!|\psi(\theta,\phi)\ket \!=\! \cos(\theta/2) |0\ket + e^{i\phi}\sin(\theta/2)|1\ket$. It can be written as function of the polar $\theta$ and azimuthal $\phi$ angles and the eigenstates $\{|0\ket, |1\ket\}$ of the Pauli matrix $\hat{\sigma}_z$. In this case, the fidelity (\ref{eq:estimativa_FRG}) depends only on the parameter $\varphi_1$ and the QSL time, as shown by

\eq{\lb{eq:fidelity_1/2}
 \begin{split}
	\xoverline{F_R}(t^R_{\star}) &= \big|\sqrt{F} \bra\psi(0)|R^{\dagger}(t^R_{\star})|\psi(0)\ket \\
    	&\hspace{0.5cm}+ \sqrt{1\!-\!F}e^{i\varphi_1} \bra\psi(0)|R^{\dagger}(t^R_{\star})|\psi^{\perp}_{1}\ket \big|^{2}.
 \end{split}
}
The perpendicular state is easily obtained by $|\psi^{\perp}_{1}\ket \!=\! |\psi (\theta\!-\!\pi,\phi) \ket$. According to the experimental setup (see Appendix \ref{sec:level1}), the axial Larmor frequency is $\omega_0 \!=\! \omega_p \!=\! 2\pi(161.975$MHz) and the radio frequency in the perpendicular direction is $\omega_1 \!=\! 2\pi(21.930$kHz), while the initial state is defined by the angles $\theta \!=\! 24.48^{\circ}$ and $\phi \!=\! 4.02^{\circ}$. For these parameters we show in Fig. \ref{fig:qsl_1/2}  the experimental fidelity of the system state in the laboratory frame (green asterisks). To obtain theses values we performed quantum state tomography at each $0.5 \;\mu$s in the time interval $[0, 22\;\mu\tn{s}]$ and used the Eq. (\ref{eq:fidelity}) to obtain the fidelity at each time. 

The system dynamics is well described by the unitary evolution governed by Hamiltonian (\ref{eq:hamiltonian_1/2}) provided that the relaxation times $T_1 \!=\! 1.96$ s and $T_2 \!=\! 170$ ms are much longer than $22\;\mu$s. In order to establish a comparison with the theoretical predictions, first we solve the Schr\"odinger equation (Appendix \ref{sec:ExactSol}). This solution is used to calculate the fidelity (\ref{eq:fidelity}) in the laboratory frame $F(t)$ (solid black line). Due to its very fast oscillating behavior, it looks like a black belt. It is possible to see in the inset of Fig. \ref{fig:qsl_1/2} the oscillating behavior of the fidelity as the system state evolves in time with the help of a time scale 1000 times shorter. Such oscillations are proportional to the ratio between the values of $\omega_0$ and $\omega_1$ and depend on the chosen initial quantum state. Indeed, the experimental values do not present oscillating behavior once the measurements are performed in a time interval that is two orders of magnitude greater than one period of oscillation of the fidelity.

\begin{figure}[t!]
 \centering
  \includegraphics[scale=0.55]{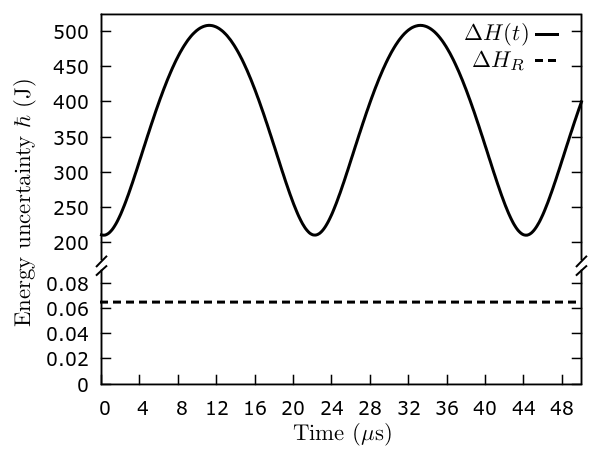}
 \caption{\small Energy uncertainty of the spin 1/2 system described by Hamiltonian (\ref{eq:hamiltonian_1/2}). The solid line is the energy uncertainty in the laboratory frame, while the dashed line is the uncertainty in the rotating frame defined by the rotation $R(t) \!=\! e^{-i \hat{\mathbf{I}}_{z}\omega_{p}t}$, i.e., a frame where the system Hamiltonian becomes time independent. The values of the physical parameters are described above in the main text.}
 \label{fig:uncertainty}
\end{figure}

Although the purpose of Ref. \cite{Deffner_2013b} was to obtain an estimate for the minimum evolution time for non-Markovian dynamics, it was shown in \cite{Mirkin_2016} that such formulas are more useful to estimate the actual evolution time. In the former, the authors make use of different norms of the generator of the evolution to get tighter bounds for the passage time. They found that the operator norm furnishes the best estimate for the actual evolution time. In  Appendix \ref{sec:qsl_comp} we compare the average time for different norms, as proposed in Ref. \cite{Deffner_2013b}, with that one obtained via AAT (\ref{eq:qsl}) and through our method  (\ref{eq:transcendental}) for spin-1/2 case. We observed that predictions made by the AAT and operator norm approaches are similar, although the former provides a tighter bound in relation to the actual evolution time. That is why we have used the AAT as the usual approach to compare with our method (blue circles). In Fig. \ref{fig:qsl_1/2},  predictions made by the AAT (\ref{eq:qsl}) in the laboratory frame are represented by a red dashed line. This last result is a consequence of the high average speed in which the system state evolves in the laboratory frame or, equivalently, it is due to its higher uncertainty energy $\Delta H (t)$, as presented in Fig. \ref{fig:uncertainty}. We emphasize that predictions of our method (blue circles) are closer to the actual evolution time, providing in some cases times estimates up to four orders of magnitude better than the AAT. If we compute the ratio $\frac{\Delta H(t)}{\Delta H_{R}}\!\sim\! 10^{4}$, then it corroborates the predictions made above. Similar observations cannot be pointed out by applying the original AAT, which one can predict in the worst case times that are four orders of magnitude less than the actual time. For short times (see inset of Fig. \ref{fig:qsl_1/2}) we observe that our method works very well, although a few roots were predicted a little bit later compared to the actual evolution time obtained theoretically. This fact is attributed to numerical errors due to the high value of the Larmor frequency $\omega_0$ in relation to the radio frequency $\omega_1$.
 
\subsection{Spin-3/2 system}
We are interested in to apply the method exposed in this work in a slightly more complex quantum system, aiming to understand how hard the application of the method becomes with the increasing of the Hilbert space dimension. The physical system is constituted by spins 3/2 of $^{23}$Na nuclei present in sodium dodecyl sulfate of a lyotropic liquid crystal evolving under the action of an external magnetic field and an internal electric field gradient \cite{Auccaise_2013}. The Hamiltonian for this system is
\eq{\lb{eq:hamiltonian_3/2}
 \begin{split} 
	\hat{H}(t) = \hbar\bigg\{ &\omega_0 \hat{\mathbf{I}}_z + \omega_1 \big[ \cos(\omega_p t)\hat{\mathbf{I}}_x + \sin(\omega_p t)\hat{\mathbf{I}}_y \big]\\
	     &+\frac{\omega_Q}{6} \big(3\hat{\mathbf{I}}^2_z- \hat{\mathbf{I}}^2 \big) \bigg\},
 \end{split}
}
where $\hat{\mathbf{I}}_j$ $(j \!=\! x, y, z)$ are the components of the nuclear spin operator and $\hat{\mathbf{I}}^2$ is the Casimir operator of the su(2) algebra. The first term is due to Zeeman interaction with frequency $\omega_0$, the second term describes the coupling with the radio-frequency field, as in the previous example, while the third one is due to the interaction of the quadrupole moments of the nuclei with the internal electric field gradient, whose coupling strength is $\omega_Q$. The strength of the Larmor frequency $\omega_0$, the radio-frequency pulse $\omega_1$, and quadrupolar coupling $\omega_Q$ are $2\pi (105.842$MHz), $2\pi (392.70$kHz), and $2\pi (15$kHz), respectively, which enable us to neglect the quadrupolar term in the Hamiltonian (Appendix \ref{sec:level1}). Additionally, we choose the precession frequency $\omega_p\!=\!\omega_0$. Therefore, in this approximation the Hamiltonian (\ref{eq:hamiltonian_3/2}) becomes quite similar to (\ref{eq:hamiltonian_1/2}), except by the fact that the spin is no longer 1/2. As the $T_{2}$ and $T_{1}$ relaxation times of the $^{23}$Na nuclear spins are 2.6$\pm$0.3 and 12.2$\pm$0.2 ms, respectively, the time evolution of the system is very well approximated by a unitary dynamics.

The initial state is a pseudo-nuclear-spin coherent state defined by the angles $\theta \!=\!0$ and $\phi \!=\! 0$, i.e., an eigenvector of $\hat{\mathbf{I}}_z$ which points out to the north pole of the Bloch sphere, $|j \!=\! 3/2, m\!=\!+3/2\ket$. For more details about the state preparation tomography of the system dynamics, see Ref. \cite{Auccaise_2013}.

\begin{figure}[t!]
 \centering
  \includegraphics[scale=0.38]{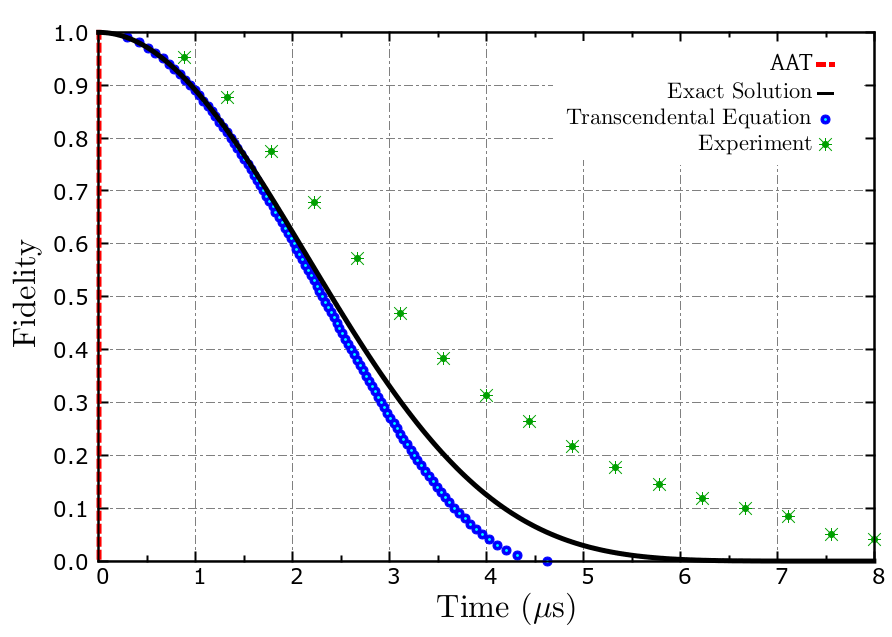}
 \caption{\small (Color online) Time estimation for the nuclear spin 3/2 of the $^{23}$Na atoms to flip from south to the north pole on the Bloch sphere. The experimental measurements are the green asterisks, whose fidelity was obtained after performing quantum state tomography at 0.5 $\mu$s. The theoretical description obtained by the exact solution of the Schr\"odinger equation (see Appendix \ref{sec:ExactSol}) is given by the solid black line, while the predictions made by the AAT and the first chronological roots of the transcendental equation are the red dashed line and blue circles, respectively. The predictions made by the transcendental equation method is two orders of magnitude better than the AAT.}
 \label{fig:qsl_3/2}
\end{figure}
Given the similarities between the spin-1/2 Hamiltonian (\ref{eq:hamiltonian_1/2}) and the spin $I\!=\!3/2$ Hamiltonian, we directly conclude that the unitary transformation that removes the time dependence of the Hamiltonian in the rotating frame is $R(t) \!=\! e^{-i \hat{\mathbf{I}}_z \omega_pt}$ and the corresponding Hamiltonian is $\hat{H}_R\!=\!\hbar\omega_1 \hat{\mathbf{I}}_x$. Then, the fidelity of the system state in the rotating frame as function of the fidelity computed in the laboratory frame and the estimated actual time can be expressed as
\eq{
 \begin{split}
     \xoverline{F}(t^R_{\star}) = &  \big|\sqrt{F} \!\bra\psi(0)|R^{\dagger}(t^R_{\star})|\psi(0)\ket \\
    & + a_1e^{i\varphi_1}\! \bra\psi(0)|R^{\dagger}(t^R_{\star})|\psi^{\perp}_{1}\ket \\
	& + a_2e^{i\varphi_2} \!\bra\psi(0)|R^{\dagger}(t^R_{\star})|\psi^{\perp}_{2}\ket \\
    & + a_3e^{i\varphi_3} \!\bra\psi(0)|R^{\dagger}(t^R_{\star})|\psi^{\perp}_{3}\ket \big|^{2},
 \end{split}
}
where $ \varphi_1, \varphi_2, \varphi_3 \!\in\! \mathds{R} $ are relative phases to the initial state. According to the normalization condition, we have that $a_1^2\!+\!a_2^2\!+\!a_3^2\!=\!1\!-\!F$, which is the equation of a sphere of radius $\sqrt{1\!-\!F}$. By virtue of this, we can parametrize $a_1,a_2,a_3 \!\in\! \mathds{R}$ by $u,v \in [0, \pi/2]$ through the relations: $a_1 \!=\! \sqrt{1\!-\!F} \sin u\cos v$, $a_2 \!=\! \sqrt{1\!-\!F} \sin u \sin v$, and $a_3 \!=\! \sqrt{1\!-\!F} \cos u$. Observe that we are restricted to one octant of the sphere, once the others ones can be recovered by choosing the correct sign of the relative phases $\varphi_1, \varphi_2, \varphi_3$. Thus, solving the transcendental equation (\ref{eq:transcendental}) numerically for each combination of parameters in the set $\{\varphi_1, \varphi_2, \varphi_3,u,v\}$ and looking for its firsts chronological roots for a given fidelity $F$, we present ours results in Fig. \ref{fig:qsl_3/2}.

By performing quantum state tomography on the nuclear spin 3/2 evolving under the action of the Hamiltonian (\ref{eq:hamiltonian_3/2}) at each discrete time $t_n \!=\! n\frac{8 \mu \text{s}}{18}$ $(n \!=\! 0, 1, \dots , 18)$ it was possible to obtain the experimental fidelity (green asteriks) with the help of Eq. (\ref{eq:fidelity}). The theoretical description (solid black line) was achieved through the exact solution of the Schr\"odinger equation for the same Hamiltonian \cite{Auccaise_2013}, resulting in the time-dependent fidelity in the laboratory frame (see Appendix \ref{sec:ExactSol})
\eq{\label{eq:fidelity_lab_3/2}
 F(t) =  \cos^6\left(\frac{\omega_{1}t}{2}\right).
}
We observe that after 8 $\mu$s the spin is totally inverted achieving the final state $|j\!=\!3/2,m\!=\!-3/2\ket$. Differently from the spin-1/2 case, we do not observe fast oscillations in the fidelity. This happens because the initial state of the system acquires only a global phase under the action of the rotation that connects the laboratory and rotating frames. Furthermore, we used Eq. (\ref{eq:fidelity_lab_3/2}) and the exact expression for the average energy uncertainty $\xoverline{\Delta H(t)}$ applied to the AAT in laboratory frame to estimate the actual evolution time (red solid line) (see also Appendix \ref{sec:ExactSol}). Similarly to the spin-1/2 case, the higher values of the average energy uncertainty (speed) in the laboratory frame produce poor predictions of the actual evolution time. On the other hand, using the predictions made by the first chronological roots of the transcendental equation (blue points), we obtain results very close to the actual evolution time. This occurs because this time estimation is performed in a rotating frame where the energy uncertainty is smaller than in the laboratory frame.

We call the attention to the fact that the dynamics of the 3/2-spin is quite similar to the NOT gate, since the initial state $|j\!=\!3/2,m\!=\!+3/2\ket$ subjected to the Hamiltonian Eq. (\ref{eq:hamiltonian_3/2}) returns after 8$\mu$s the state $|j\!=\!3/2,m\!=\!-3/2\ket$. Therefore, we are also estimating the running time of a quantum gate.

\section{Discussions and perspectives}\lb{sec:discussions}
In this work we proposed a method to estimate the actual evolution time of a quantum system based on a modified version of the quantum-speed-limit time evaluated in a reference frame where the system state evolves as close as possible to the geodesic path over the underlying Hilbert space. Such a method enabled us to obtain a good time estimation for given 1/2 and 3/2 nuclear spin dynamics, obtaining values up to four orders of magnitude better than the AAT. There are two important issues related to the implementation of our method: The first one is the choice of a reference frame where the speed of the state evolution (average energy uncertainty) is smaller than in the laboratory frame and the second one is the computational cost to implement it. Related to the first issue, our strategy was to find a reference frame where the Hamiltonian becomes time independent. This is interesting because it is possible to evaluate exactly the energy uncertainty using only the initial state and the Hamiltonian at the rotating frame so that no approximation at this point of the calculation is necessary.

Furthermore, insofar as in NMR systems it is always possible to go to a rotating frame defined by the time-dependent radio-frequency field \cite{Abraham_1994}, the remaining Hamiltonian becomes time independent or varies slowly in time. Related to the computational cost to solve the transcendental equation (\ref{eq:transcendental}) for the shortest time, there are efficient methods to solve this kind of equation \cite{Luck_2015}, although we have used the bisection method. According to this method, the number of iterations to converge to a root to within a certain level of tolerance (error) depends on the value of $t_{\star}^R$ in the following way: $\log t_{\star}^R $. Furthermore, for a $n$-dimensional Hilbert space, we need to solve $2(n\!-\!1) N^2$ transcendental equations, where $N$ is the number of steps to sweep over each coefficient in state (\ref{eq:state_expansion}). After that we take the first chronological root (time) among all solutions. Therefore, although this method can provide good results, it can become expensive for higher dimensional systems, scaling as $2(n\!-\!1) N^2 \log t_{\star}^R$. On the other hand, the actual evolution time can be found through the solution of the time-dependent Schr\"odinger equation. There are several methods to solve such equation numerically \cite{Kosloff_1988}. For example, if we are considering the second-order differencing scheme, then the computational cost is $O\left( t_{\star}^R \right)^{3/2}$ \cite{Kosloff_1988,Leforestier_1991}.

For future applications of our method, we would like to diminish its computational cost.  Another possibility to improve it would be to find the unitary transformation that forces the quantum state to evolve over a geodesic length in the new reference frame.
\section{Acknowledgements} 

The authors thank to R. M. Filho, M. M. Taddei, L. C. Celeri, P. H. S. Ribeiro, D. Soares-Pinto, F. Toscano, and L. G. C. Rego for valuable discussions, the anonymous referees for relevant comments, A. S. Magno for reading the manuscript, and also the Brazilian National Institute for Science and Technology of Quantum Information and the Brazilian funding agencies CNPq and CAPES for the financial support.

\appendix
\section{Replacement of $F_{R}(t)$ by $\xoverline{F}_{R}(t_{\star}^R)$} \label{app:error_fidelity}
Our method consists in to solve the equation
\begin{equation}\label{eq:qslRotating_app}
    \xoverline{\Delta H_R}t_{\star}^R = \hbar \arccos \sqrt{F_R(t)},
\end{equation}
i.e., to find the minimum time $t_{\star}^R$ which is solution of the above equation. As the fidelity in the laboratory frame $F$ is an information given \emph{a priori}, to connect it to the fidelity in the rotating frame $F_R(t)$ it is necessary to write the former as $F= |\bra \psi(0)|\psi(t)\ket|^2$ so that $F_R(t)=|\bra \psi(0)|R^{\dagger}(t)|\psi(t) \ket |^2 $. Once $|\psi(t)\ket$ is unknown, we can replace it by
\begin{equation} \label{eq:state_expansion_app}
 |\xoverline{\psi} (t_{\star}^R) \ket = \sqrt{F}|\psi(0)\ket + \sum^{n-1}_{j=1}a_{j}e^{i\varphi_{j}}|\psi^{\perp}_{j}\ket,
\end{equation}
since it preserves the original value of the fidelity. However, the fidelity $F_R(t)$ keeps its dependence on the actual evolution time through $R^{\dagger}(t)$. Such a fact makes it impossible to find the solution of Eq. (\ref{eq:qslRotating_app}). To overcome this obstacle we introduced the identity operator $R^{\dagger}(t^R_{\star})R(t^R_{\star})=\bold{1}$ in the expression for $F_R(t)$,
\begin{equation} \label{fidelity_app_1}
F_R(t)=|\bra \psi(0)|R^{\dagger}(t^R_{\star})\left[ R(t^R_{\star})R^{\dagger}(t) \right]|\xoverline{\psi} (t_{\star}^R) \ket |^2 ,
\end{equation}
and analyze the term in brackets. The unitary transformations we are considering are of the form $R(s)=e^{-i\Lambda s/\hbar}$, with $\Lambda = \Lambda^{\dagger}$ and $s \in \mathds{R}$. With the help of Eq. (\ref{times}), we obtain
\[
R(t^R_{\star})R^{\dagger}(t)=e^{-i\frac{\Lambda}{\hbar}\left( t_{\star}^R -t \right)}=\tn{exp} \left[ -i\frac{\Lambda t}{\hbar}\left(\frac{\ell_{\tn{geodes}}^R}{\ell_{\tn{real}}^R} -1 \right) \right]. 
\]
From the equation above we observe that for $\ell_{\tn{geodes}}^R \rightarrow \ell_{\tn{real}}^R$, $R(t^R_{\star})R^{\dagger}(t) \rightarrow \bold{1}$ and the replacement of $F_{R}(t)$ by $\xoverline{F}_{R}(t_{\star}^R)$ becomes a good approximation. Then, according to our method, the useful expression is
\begin{equation} \label{eq:transeq}
    \xoverline{\Delta H_R}t_{\star}^R = \hbar \arccos \sqrt{\xoverline{F}_R(t_{\star}^R )},
\end{equation}
where $\xoverline{F}_R(t_{\star}^R)=|\bra \psi(0)|R^{\dagger}(t_{\star}^R)|\xoverline{\psi} (t_{\star}^R) \ket |^2$ and the errors in our approach are dependent on the ratio between the lengths of the geodesic and actual paths in the rotating frame, as expected.

\section{Exact solution of the Schr\"odinger equation}\label{sec:ExactSol}
Our goal here is to solve the Schr\"odinger equation
\begin{equation}\label{eq:SchrodingerApp}
i \hbar \frac{\partial | \psi (t) \ket}{\partial t} = \hat{H}(t) | \psi (t) \ket,
\end{equation}
for the 1/2-spin and spin-3/2 spins dynamics and to obtain the time-dependent fidelity $F(t)$ in order to compare with the predictions of our method and experimental data.   
\subsection{Spin-1/2 system}
The spin-1/2 Hamiltonian is
\begin{equation} \label{eq:HamiltonianApp}
  \hat{H}(t) =  \hbar \left\{ \omega_{0} \hat{\mathbf{I}}_{z} + \omega_{1} \big[ \cos(\omega_{p} t) \hat{\mathbf{I}}_{x} + \sin(\omega_{p} t)\hat{\mathbf{I}}_{y} \big] \right\}.
\end{equation} 
First we apply the unitary transformation $R(t)\!=\!e^{-i\hat{\mathbf{I}}_z\omega_pt}$  on Eq. (\ref{eq:SchrodingerApp}), obtaining the transformed Hamiltonian
\eq{\lb{eq:TransHamiltonianApp}
 \begin{split}
  \hat{H}_R &= R^{\dagger}(t)\hat{H}(t)R(t) -i\hbar R^{\dagger}(t)\dt{R}(t), \\
	&= \hbar \omega_1 \hat{\mathbf{I}}_x,
 \end{split}
} 
where we set $\omega_p\!=\!\omega_0$. Then, the evolved state in this new frame is 
\begin{equation} \label{eq:stateGirApp}
	|\psi_R (t) \ket = e^{-i \hat{\mathbf{I}}_x\omega_1 t } |\psi_R (0)\ket, \quad \big( |\psi_R (0)\ket=|\psi(0)\ket \big).
\end{equation}
Coming back to the original picture, we obtain the exact solution of the Schr\"odinger equation as
\begin{equation}
	|\psi(t) \ket = e^{-i \hat{\mathbf{I}}_z \omega_p t}e^{-i \hat{\mathbf{I}}_x \omega_1 t} |\psi(0) \ket.
\end{equation}
A general pure state on the Bloch sphere can be written as $|\psi(0)\ket \!=\! \cos\!\left( \theta/2 \right)|0\ket \!+\! e^{i\phi} \sin\!\left(\theta/2 \right)|1\ket$, with $\theta$ and $\phi$ defining the polar and azimuthal angles, respectively. Therefore, the evolved state at time $t$ is represented in matrix form as
\begin{equation} \label{stateTwo-levelApp}
   \hspace{-0.1cm}|\psi(t) \ket \!= \!\left(\!\begin{matrix}
e^{-i\omega_0 t /2}\!\left[ e^{-i\Omega t/2}c_{+}\!\cos(\chi/2)\!+\!e^{i\Omega t/2}c_{-}\!\sin(\chi/2) \right] \\
e^{i\omega_0 t /2}\!\left[ e^{-i\Omega t/2}c_{+}\!\sin(\chi/2)\!-\!e^{i\Omega t/2}c_{-}\!\cos(\chi/2) \right]
\end{matrix}
 \!\!\right)\!,
\end{equation}  
with $\Omega\!=\!\sqrt{\omega_0^2\!+\!(\omega_1\!-\!\omega_0)^2}$, $\cos \chi \!=\! (\omega_1\!-\!\omega_0)/\Omega$, $\sin \chi \!=\!\omega_0/\Omega$ and,
\eq{
\begin{split}
c_{+} &= \left( \cos(\chi/2)\cos(\theta/2)+e^{i\phi}\sin{\chi/2}\sin{\theta/2} \right), \\
c_{-} &= \left( \sin(\chi/2)\cos(\theta/2)-e^{i\phi}\cos{\chi/2}\sin{\theta/2} \right).
\end{split}
}
Evaluating the fidelity, we get
\begin{widetext} 
\begin{align} 
F(t)&=\left[ \cos \left(\frac{\omega_0 t}{2} \right)\cos \left(\frac{\Omega t}{2} \right) - \cos \chi \sin \left(\frac{\omega_0 t}{2} \right)\sin \left(\frac{\Omega t}{2} \right) \right]^2 \nonumber\\
&\;\;+\Biggl\{ \left( \cos \theta \cos \chi + \cos \varphi \sin \theta \sin \chi \right) \left[ \cos \left(\frac{\omega_0 t}{2} \right)\sin \left(\frac{\Omega t}{2} \right) - \cos \chi \sin \left(\frac{\omega_0 t}{2} \right)\cos \left(\frac{\Omega t}{2} \right) \right] \nonumber\\
&\;\;+ \sin \left(\frac{\omega_0 t}{2} \right)\sin \chi \left[  \cos \left(\frac{\omega t}{2} \right)\left( \cos \theta \sin \chi - \cos \varphi \sin \theta \cos \chi \right) +  \sin \left(\frac{\omega t}{2} \right)\sin \theta \sin \varphi   \right] \Biggr\}^2. \label{eq:fidelityApp}
\end{align}
\end{widetext}
Now we are able to evaluate the energy uncertainty in both frames of reference. Using the Hamiltonian (\ref{eq:TransHamiltonianApp}) and the quantum state (\ref{eq:stateGirApp}), we find that
\eq{
	\Delta H_R = \frac{\hbar \Omega}{2}\sqrt{1-\left( \cos \chi \cos \theta + \cos \phi \sin \chi \sin \theta \right)^2},
}
while the energy uncertainty in the laboratory frame is obtained through Eqs. (\ref{eq:HamiltonianApp}) and (\ref{stateTwo-levelApp}), i.e.,
\eq{ \label{eq:uncertainty_1/2_app}
	\Delta H(t) = \sqrt{ \frac{\hbar^2}{4} \left(\omega_0^2\!+\!\omega_1^2 \right) - \bra \psi(t)|\hat{H}(t)|\psi(t)\ket^2 },
}
where
\eq{
 \begin{split}
	\bra \psi(t)|\hat{H}(t)|\psi(t)\ket =& \frac{\hbar}{2}\left( |c_+|^2-|c_-|^2\right)\left( \Omega+\omega_0\cos \chi \right)\\
	&+\hbar \omega_0 \sin \chi \big[ \cos \left( \Omega t\right) \mathbf{Re}(c_+^*c_{-})\\
    &-\sin \left( \Omega t\right) \mathbf{Im}(c_{+}^*c_{-})  \big],
\end{split}
}
with
\eq{
|c_+|^2-|c_-|^2=\cos \theta \cos \chi+\cos \phi \sin \theta \sin \chi,
}
and
\begin{subequations}
 \begin{align}
	\mathbf{Re}(c_+^*c_)&=\frac{1}{2}\left(\cos \theta \sin \chi-\cos \phi \sin \theta \cos \chi \right),\\
	\mathbf{Im}(c_+^*c_)&=-\frac{1}{2} \sin \theta \sin \phi .
 \end{align}
\end{subequations}
The AA bound in the laboratory frame depends only on the fidelity (\ref{eq:fidelityApp}) and the time average of the energy uncertainty (\ref{eq:uncertainty_1/2_app}). Such bound is evaluated numerically and is plotted in Fig. {\ref{fig:qsl_1/2} as the red dashed line.
\subsection{Spin-3/2 system}
This system was already studied in Ref.\cite{Auccaise_2013}. A particular case in which $\omega_0, \omega_1 \gg \omega_Q$ and $\omega_p\!=\!\omega_0$, the Hamiltonian in the Schr\"odinger picture (\ref{eq:hamiltonian_3/2}) simplifies to
\begin{equation}\label{eq:hamiltonian_3/2_app}
\hat{H} (t)= \hbar  \left\lbrace \omega_{0} \hat{\mathbf{I}}_{z} + \omega_{1} \left[ \cos(\omega_{0}t)\hat{\mathbf{I}}_{x} + \sin(\omega_{0}t)\hat{\mathbf{I}}_{y} \right] \right\rbrace . \nonumber
\end{equation} 
Going to the rotating frame through the the unitary transformation $R(t)\!=\!e^{-i \hat{\mathbf{I}}_z \omega_0 t }$, the new Hamiltonian is 
\eq{
\hat{H}_R = \hbar \omega_1 \hat{\mathbf{I}}_x,
}
and the evolved state in the rotating frame is 
\eq{ \label{eq:stateGir}
|\psi_R(t)\ket = e^{-i\omega_1 t \hat{\mathbf{I}}_x} |\psi(0) \ket.
}
Immediately we obtain the general expression for the evolved state in the laboratory frame as 
\begin{equation}
| \psi(t) \ket = e^{-i\omega_0 t \hat{\mathbf{I}}_z  }e^{-i\omega_1 t \hat{\mathbf{I}}_x} |\psi(0) \ket.
\end{equation}
By choosing the initial state as a pseudo-nuclear-spin coherent state characterized by the angles $\left( \theta\!=\!0, \phi\!=\!0 \right)$ \cite{Auccaise_2013}, or simply $|j\!=\!3/2,m\!=\!+3/2 \ket$, the system state at time $t$ in the laboratory frame is
\begin{equation} \label{eq:state_3/2_evolved}
   |\psi(t)\ket = \left(\!\!\begin{array}{rl}  
        ie^{-i3\omega_{0}t/2}  &    \sin^3\left(\omega_{1}t/2\right) \\
  -\sqrt{3}e^{-i \omega_{0}t/2}  &    \sin^2\left(\omega_{1}t/2\right)\cos\left(\omega_{1}t/2\right) \\
        -i\sqrt{3}e^{ i \omega_{0}t/2}  &   \sin\left(\omega_{1}t/2\right)\cos^2\left(\omega_{1}t/2\right)\\
          e^{ i3\omega_{0}t/2}  &   \cos^3\left(\omega_{1}t/2\right)  
   \end{array}\!\!\right).
\end{equation}
Consequently, the fidelity becomes
\eq{\label{eq:fidelity_3/2_app}
 F(t) = \cos^{6}\!\left(\frac{\omega_{1}t}{2}\right). 
}
Its behavior is represented by the solid black line in Fig. \ref{fig:qsl_3/2}. To calculate the time estimation it is necessary the energy uncertainty in the rotating frame, which reads 
\eq{
\Delta H_R = \frac{\sqrt{3}}{2}\hbar \omega_0,
}
while in the laboratory frame it is obtained through the substitution of (\ref{eq:hamiltonian_3/2_app}) and (\ref{eq:state_3/2_evolved}) in  the expression $\Delta H(t)\!=\!\sqrt{\bra \psi (t)|\hat{H}^2(t)|\psi (t)\ket \!-\! \bra \psi (t)|\hat{H}(t)|\psi (t)\ket^2}$. Due to its long length, the expressions for $\Delta H(t)$ and $\xoverline{\Delta  H(t)}$ were evaluated numerically. Therefore, by combining $\xoverline{\Delta  H(t)}$ and (\ref{eq:fidelity_3/2_app}) according to Eq.(\ref{eq:qsl}), we obtain the AA bound for the spin-3/2 system, which is reported in Fig. \ref{fig:qsl_3/2} by the dashed red line.   

\section{Comparing average times of the state evolution}\lb{sec:qsl_comp}
The formulas below were deduced in Ref. \cite{Deffner_2013b} for QSL time for unitary dynamics, although it was shown later in Ref. \cite{Mirkin_2016} that they are more appropriate to estimate the actual evolution time:  
\begin{figure}[t!]
 \centering
  \includegraphics[width=0.48 \textwidth]{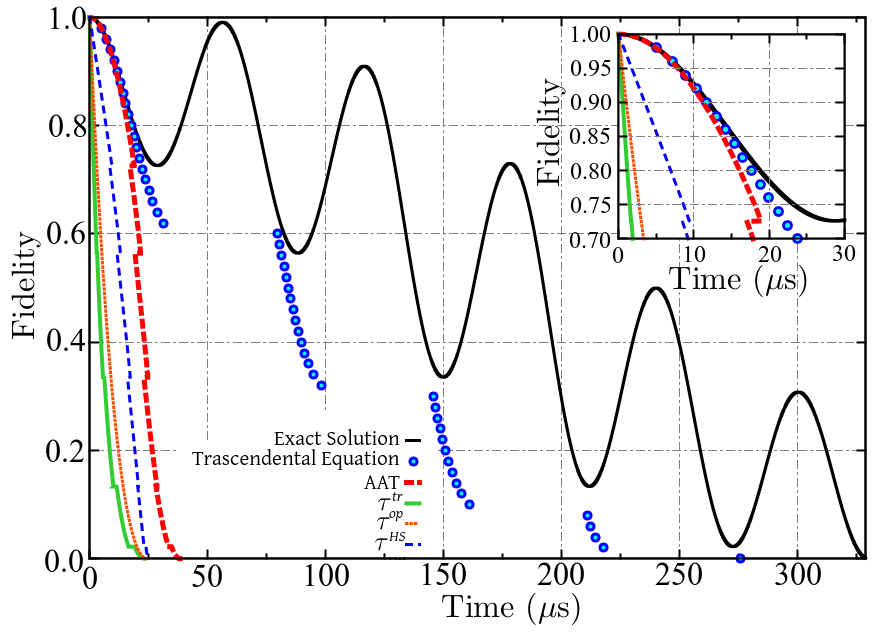}
 \caption{(Color online) Predictions of the actual evolution time for a spin 1/2 system evolving according to the Hamiltonian (\ref{eq:HamiltonianApp}). The initial Bloch state is characterized by the angles $\theta\!=\!30^{\circ}$ and $\phi\!=\!180^{\circ}$ and the frequencies used to obtain these plots are $\omega_0=2\pi (16,000$Hz), $\omega_1=2\pi (1,250$Hz), and $\omega_p=2\pi (15,278$kHz). The black solid line describes the time-dependent fidelity obtained through the exact solution of the Schr\"odinger, which one will be used as reference. The predictions made using the trace norm (thin green solid line) in the QSL time expression are the worst ones. By using the operator norm (orange dotted line) and Hilbert-Schmidt norm (blue thin dashed line), we observe an improvement on the time estimate, although the AA bound (red dashed line) furnishes better results when compared to the previous ones. However, the predictions made by the transcendental equation method (blue circles) are closer to the actual time. In the inset such hierarchy among the QSL times is also preserved in a short time scale.}
 \label{fig:spin1/2_appendix}
\end{figure}

\begin{equation} \lb{eq:qsl_norm}
\tau^{\tn{tr,op}} = \frac{\hbar \sin^2\left( \mathcal{L}\right)}{2\Lambda_{\tau}^{\tn{tr,op}}}, \qquad \tau^{\tn{HS}} = \frac{\hbar \sin^2\left( \mathcal{L}\right)}{\Lambda_{\tau}^{\tn{HS}}},
\end{equation}
where $\mathcal{L}\!=\!\arccos \big[ \sqrt{F(t)} \big]$ is the geodesic length between the initial and final states and
\[
\Lambda_{t}^{\ell} = \frac{1}{t}\int^{t}_0  \norm{H(t)|\psi(t^{'})\ket \bra \psi(t^{'})| }_{\ell}dt^{'}.
\]
The index $\ell$ refers to the type of norm being used. For operator norm 
\begin{equation} \lb{eq:op}
   \norm{\hat{H}(t)|\psi(t^{'})\ket \bra \psi(t^{'})| }_{\text{op}}=\sigma_1,
\end{equation}
for trace norm
\begin{equation} \lb{eq:tr}
   \norm{\hat{H}(t)|\psi(t^{'})\ket \bra \psi(t^{'})| }_{\text{tr}}= \sum_{i=1}^n\sigma_i,
\end{equation}
and for the Hilbert-Schmidt norm
\begin{equation} \lb{eq:HS}
   \norm{\hat{H}(t^{'})|\psi(t^{'})\ket \bra \psi(t^{'})| }_{\text{HS}} = \sum_{i=1}^n\sigma_i^2.
\end{equation}
The $\sigma_i$ are singular values of $\hat{H}(t^{'})|\psi(t^{'})\ket \bra \psi(t^{'})|$ and are written in descending order as $\sigma_1 \!\geqslant\! \sigma_2 \!\geqslant\! \dots \!\geqslant\! \sigma_n$. In Fig. \ref{fig:spin1/2_appendix} we compare the expressions given by Eq. (\ref{eq:qsl_norm}) with that ones from the AA bound (\ref{eq:qsl}) and transcendental equation (\ref{eq:transcendental}) for the example of the spin-1/2 system. Despite the hierarchic relation $\Lambda^{\tn{op}} \!\leqslant\! \Lambda^{\tn{HS}} \!\leqslant\! \Lambda^{\tn{tr}}$ \cite{Deffner_2013b}, in Fig. \ref{fig:spin1/2_appendix} we noticed that $\tau^{\tn{HS}}$ is bigger than the other two expressions obtained in \cite{Deffner_2013b}, but smaller than the AA bound because of the inequality $\mathcal{L} \!\geq\! \sin^2{\mathcal{L}}$. Altogether $\tau^{\tn{tr}}$ (green thin solid line), $\tau^{\tn{op}}$ (orange dotted line), $\tau^{\tn{HS}}$ (blue thin dashed line), and the AA bound (\ref{eq:qsl})  (red thick dashed line) furnish similar results for the time estimates. Nevertheless, the best result is given by the solution of the transcendental equation (\ref{eq:transcendental}) (blue circles). Therefore, for the sake of simplicity, hereafter we will compare the predictions for the actual time made by AA bound (\ref{eq:qsl}), first roots of transcendental equation (\ref{eq:transcendental}), and experimental data.

\section{Experimental data of the spin system $I=1/2$} \lb{sec:level1}
The NMR implementation is carried out using an Ascend Bruker of 400-MHz spectrometer (Larmor frequency of $^{1}$H nuclei) at State University of Ponta Grossa. A double-channel probe head for liquid samples is used to apply a transverse magnetic field of a few gauss. The channel configuration obeys the label (H/F)X, which means that the first channel is dedicated to detecting $^{1}$H and $^{19}$F nuclei signals and the second channel detects signals of many nuclear species between 162 MHz and 40 MHz, such as $^{31}$P, $^{13}$C, $^{23}$Na, and $^{15}$N. 

The solution NMR experiment is performed using molecules of o-phosphoric acid (H$_{3}$PO$_{4}$) dissolved in deuterated water (D$_{2}$O), at room temperature   {$(\sim 25^{\circ}$C)}. The stoichiometry of our sample obeys 12.5 \% of H$_{3}$PO$_{4}$ and 87.5 \% of D$_{2}$O, a solution of 650 ml was placed in a glass of 5-mm NMR tube. Sketched in Fig. \ref{fig:NMRscheme}(a) is a picture of H$_{3}$PO$_{4}$ molecular structure. The choice of H$_{3}$PO$_{4}$ molecule to achieve an isolated one-qubit system using $^{31}$P nuclei obeys some chemical characteristics.  First, the $^{31}$P isotope is naturally abundant at 100\%. Second, oxygen nuclei has zero nuclear spin, consequently there is not any coupling strength between oxygen nuclei and $^{31}$P. Third, the electrons of the four oxygen nuclei are arranged in such a away that they shell the interaction between $^{1}$H nuclei and $^{31}$P nuclei. Fourth,  to verify the efficiency of this shielding process, in Fig. \ref{fig:MagnetizationTomography}c we present the spectrum of $^{31}$P. The peak width at half height is $\thicksim 2.1$ Hz, meaning that if there exist any kind of coupling between any neighborhood nuclei of  $^{31}$P, then the strength of the interaction is less than 2.1 Hz, which is considered weak.  The calibration parameters on the $X$ channel of the probe-head in to apply and to detect  $^{31}$P nuclei signals are: tuning of the radio-frequency is  $\omega_{p} =2 \pi \left( 161.975 \ \text{\tn{MHz}}\right)$, $\pi/2$ pulse time calibrated at $11 \ \mu$s which corresponds to a strength radio-frequency of $\omega_{1}=2 \pi \left( 21.93 \ \text{\tn{kHz}}\right)$, recycle delay  $d_{1}= 20$s, acquisition time  $\tau_{Acq.}=2.5$ s.

At room temperature, the theoretical formalism to describe the quantum state of any nuclear spin system is the maximum mixture state\cite{Oliveira_2007} 
\begin{equation}
\hat{\rho}\approx \frac{1}{{\mathcal{Z}}}\hat{\mathbf{1}}+\frac{\beta {\hbar
\omega _{0}}}{{\mathcal{Z}}}\hat{\mathbf{I}}_{z}\text{,}  \label{densitymatrix}
\end{equation}
where $\beta =1/k_{B}T$, ${\mathcal{Z}}={\tn{Tr}\left[ e^{\left( -\beta \hbar
\omega _{0}\hat{\mathbf{I}}_{z}\right) }\right] }$ is partition function, $T$ is temperature, $k_{B}$ is Boltzmann's constant, ${\hbar }$ is reduced Planck's constant and ${\omega _{0}=2\pi }\left( 161.975\ \text{\tn{MHz}}\right) $ is Larmor frequency for $^{31}$P nuclei at 9.39 Tesla. Thus, the polarization factor is $\epsilon =\frac{\beta {\hbar \omega _{L}}}{2{\mathcal{Z}}} = 0.652\times 10^{-5}$, which is a slight deviation from the normalized identity matrix $\hat{\mathbf{1}}/\mathcal{Z}$. In this sense, the thermal state of Eq. (\ref{densitymatrix}) can be rewritten
\begin{equation}
\hat{\rho}\approx \left( \frac{1}{\mathcal{Z}} -\epsilon \right)\hat{\mathbf{1}}+\epsilon \hat{\rho}_{0},  \label{pseudopurestate}
\end{equation}%
where $\hat{\rho}_{0}$ represents the pure part of density matrix with unitary trace.

\begin{figure}[t!]
 \centering
  \includegraphics[width=0.48 \textwidth]{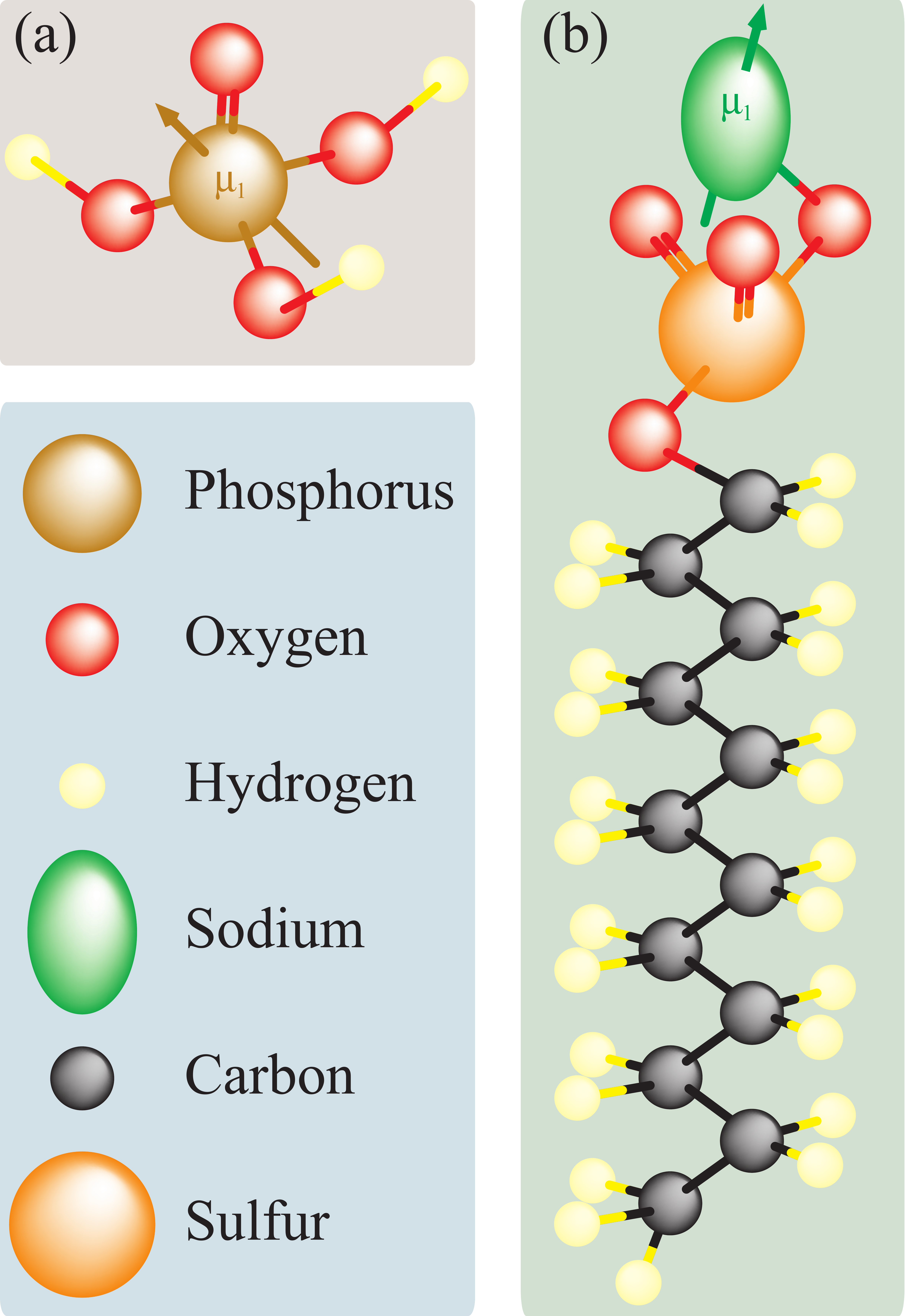}
 \caption{(Color online) We sketch a cartoon of the molecules used in the present study. (a) O-phosphoric acid molecule (H$_{3}$PO$_{4}$). (b) Sodium dodecil sulfate (Na-C$_{12}$H$_{25}$SO$_{4}$).}
 \label{fig:NMRscheme}
\end{figure}

\subsection{The quantum state tomography procedure}\lb{sec:level1C}
Tomography was performed of the pure part of the density matrix, which was reconstructed using global rotations \cite{Teles_2007}. In this sense, the NMR technique detects magnetization along the  $x$-axis and $y$-axis which correspond to first order coherences of any density matrix, and this characteristic has been explored into detect other orders of coherences. Therefore, we present a brief explanation for spin  $I=1/2$, and analogous procedure is extended for spin $3/2$, see details in \cite{Teles_2007,Auccaise_2013}.

In order to apply the tomography procedure on a spin system  $I=1/2$, we identify zero- and first-order coherences on its density matrix. Thus, we must use appropriate rotations to transfer intensities from zeroth-order of coherences into first-order coherences. The density operator for one qubit is represented by $2\times 2$ operator. Therefore, as an example, lets us reconstruct the density matrix of the nuclear spin in its ground state, $\left\vert \uparrow \right\rangle$. So, we represent the elements of any density matrix as%
\begin{equation}
\hat{\rho}_{0} =\left[ 
\begin{array}{cc}
\rho_{\left\vert  \uparrow \right\rangle   \left\langle \uparrow \right\vert} & \rho_{\left\vert  \uparrow \right\rangle   \left\langle \downarrow \right\vert} \\ 
\rho_{\left\vert \downarrow \right\rangle  \left\langle \uparrow \right\vert} & \rho_{\left\vert \downarrow \right\rangle  \left\langle \downarrow \right\vert}%
\end{array}%
\right] =\left[ 
\begin{array}{cc}
x_{1} & x_{2}+ix_{3} \\ 
x_{2}-ix_{3} & x_{4}%
\end{array}%
\right] \text{,} \label{Densitymatrix2x2} 
\end{equation}%
and to compute numerical values for their elements, we proceed with a protocol that can be summarized into three stages:

\begin{figure}[t!]
\includegraphics[width=0.48 \textwidth]{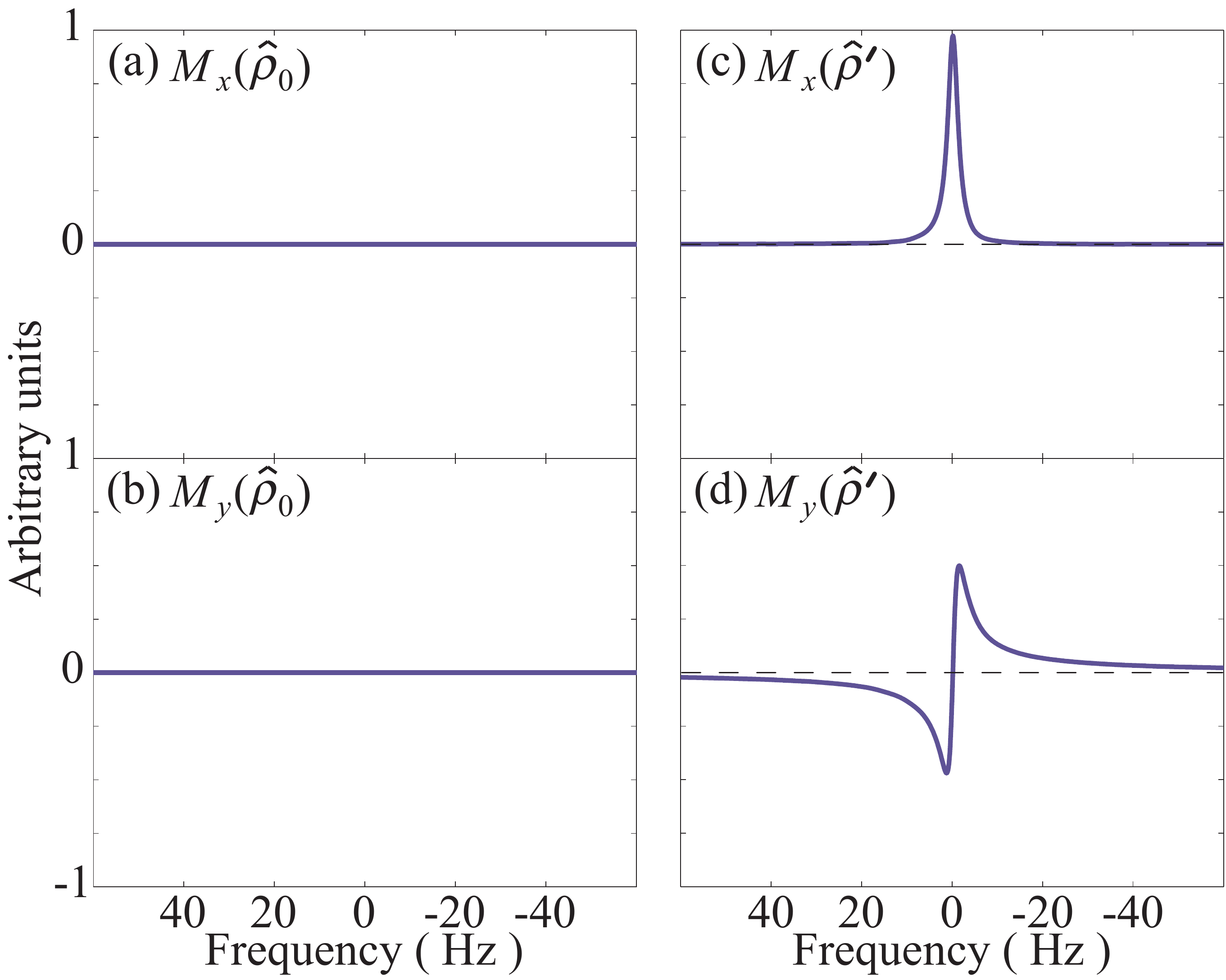}
\caption{(Color online) NMR spectra of $^{31}$P nuclei in order to detect the initial quantum state $ \left\vert \uparrow \right\rangle $ represented by the density matrix $ \hat{\rho}_{0}$. A spectral window of 120 Hz is established to show the magnetization: (a) along $x$-axis with $M_{x}\left( \hat{\rho}_{0}\right)$, (b) along $y$-axis, with  $M_{y}\left( \hat{\rho}_{0}\right)$, (c) along $x$-axis, with   $M_{x}\left( \hat{\rho}^{\prime}\right)$ such that   $\hat{\rho}^{\prime}  = R_{y}\left( \frac{\protect\pi }{2}\right)\hat{\rho}_{0}R_{y}^{\dagger}\left( \frac{\protect\pi }{2}\right)$, (d)  along $y$-axis, with   $M_{y}\left( \hat{\rho}^{\prime}\right)$. }
\label{fig:MagnetizationTomography}
\end{figure}

\textit{The first stage:} NMR observables are spin angular momentum operators $ \hat{\mathbf{I}}_{x}$ and $\hat{\mathbf{I}}_{y}$, such that their average values correspond to magnetization along the $x$- and $y$-axis, respectively. By simple calculation for any density matrix $ \hat{\rho}_{0}$, see Eq. (\ref{Densitymatrix2x2}), it is possible to show that
\begin{subequations}
\begin{align}
M_{x}\left( \hat{\rho}_{0}\right) &=\text{\texttt{Tr}}\left\{ \hat{\mathbf{I%
}}_{x}\hat{\rho}_{0}\right\} =x_{2}\text{,} \label{MagnetizationXRho} \\
M_{y}\left( \hat{\rho}_{0}\right) &=\text{\texttt{Tr}}\left\{ \hat{\mathbf{I%
}}_{y}\hat{\rho}_{0}\right\} =-x_{3}\text{,} \label{MagnetizationYRho}
\end{align}
\end{subequations}
bring information on the real and imaginary parts of the element $\rho_{\left\vert \uparrow \right\rangle \left\langle \downarrow \right\vert}$ and also its complex conjugate  $\rho_{\left\vert \downarrow \right\rangle \left\langle \uparrow \right\vert}$.

\textit{The second stage:} We transform $\hat{\rho}_{0}$ applying the operator $
R_{y}\left( \frac{\pi }{2}\right) $ and each element of the transformed density matrix can be represented by
\begin{equation}
\hat{\rho}^{\prime}=\frac{1}{2}\left[ 
\begin{array}{cc}
x_{1}+x_{4}-2x_{2} & x_{1}-x_{4}+2ix_{3} \\ 
x_{1}-x_{4}-2ix_{3} & x_{1}+x_{4}+2x_{2}%
\end{array}%
\right] \text{.} \label{Densitymatrix2x2Ry} 
\end{equation}
The magnetization along the $x$- and $y $-axes, which correspond to real and imaginary parts of the element $\rho^{\prime}_{\left\vert \uparrow \right\rangle \left\langle \downarrow \right\vert}$, are 
\begin{subequations}
\begin{align}
M_{x}\left( \hat{\rho}^{\prime}\right) &=\text{\texttt{Tr}}\left\{ \hat{\mathbf{I}}_{x}%
\hat{\rho}^{\prime}\right\} =\frac{x_{1}-x_{4}}{2} \text{,} \label{MagnetizationXRhoRodaY}\\
M_{y}\left( \hat{\rho}^{\prime}\right) &=\text{\texttt{Tr}}\left\{ \hat{\mathbf{I}}_{y}%
\hat{\rho}^{\prime}\right\} =x_{3}\text{.} \label{MagnetizationYRhoRodaY}
\end{align}
\end{subequations}
Furthermore, we use the normalization condition  of the density matrix operator, 
\begin{equation*}
1=x_{1}+x_{4}\text{.}
\end{equation*}
Therefore, the set of equations generated by the above procedure is complete.

\begin{figure}[b]
\includegraphics[width=0.48 \textwidth]{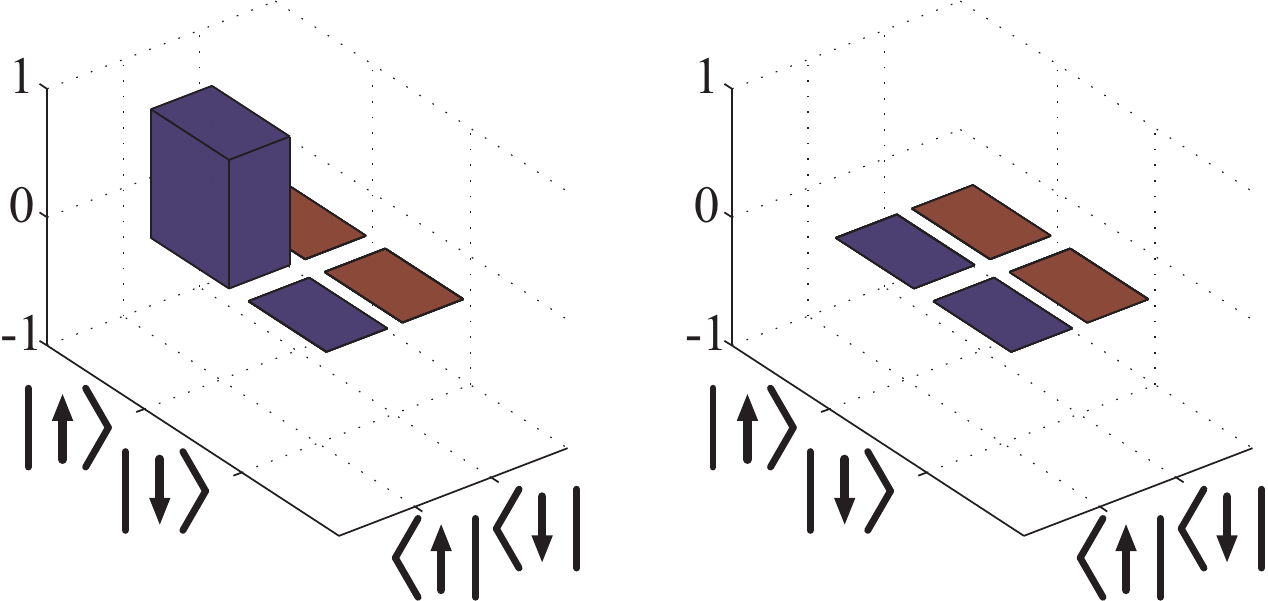}
\caption{(Color online) Tomographed quantum state. The quantum state $\hat{%
\protect\rho}=\left\vert \uparrow \right\rangle \left\langle \uparrow
\right\vert $ is showed at its matrix representation using bar charts. Left (Right)
bar chart corresponds to real (imaginay) elements.}
\label{fig:Rho21-00}
\end{figure}

\begin{figure}[t!]
 \centering
  \includegraphics[scale=0.30]{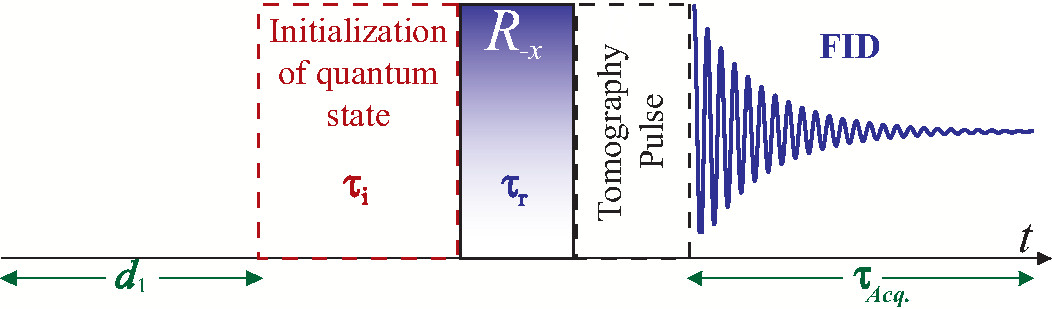}
 \caption{(Color online) NMR Pulse sequence to implement the register of the quantum dynamics protocol. First, recycle delay $d_{1}$ allows to achieve the thermal equilibrium state of all nuclei in the sample. Second, the procedure to initialize the quantum state is depicted by the first box (dashed line) of the pulse sequence with  $\tau_{\textbf{i}}$ the required time to perform the initialization of the quantum state. Third, the rotation  $R_{-x}\left( \tau_{\textbf{r}}\right)$ allow us to control the dynamics of the nuclear spin system. Fourth, the quantum state tomography procedure performing any rotation of the protocol \cite{Teles_2007} is represented by the black dashed line box. Finally, the read out of the free induction decay (FID) is performed under $\tau_{Acq}$ the acquisition time.}
 \label{fig:PulseSequence}
\end{figure}

\textit{The third stage:} In order to reconstruct the density matrix, we  need to identify each equation of the previous stages with those four spectra on Fig. \ref{fig:MagnetizationTomography}. In this sense, the spectra in Fig. \ref{fig:MagnetizationTomography}a and \ref{fig:MagnetizationTomography}b represent magnetizations $M_{x}\left( \hat{\rho}_{0}\right)$ and  $M_{y}\left( \hat{\rho}_{0}\right)$ which are quantified by Eq. (\ref{MagnetizationXRho}) and  (\ref{MagnetizationYRho}). Similarly, the spectra in Fig. \ref{fig:MagnetizationTomography}c and  \ref{fig:MagnetizationTomography}d represent magnetizations, after a $\frac{\pi}{2}$-rotation, which generate from Eqs. (\ref{MagnetizationXRhoRodaY}) and Eq. (\ref{MagnetizationYRhoRodaY}) information on the state $\rho^{\prime}$. In this procedure, we use and apply a normalization factor 2, because the amplitude of magnetization goes from -1 to 1, see Fig. \ref{fig:MagnetizationTomography}, such that the magnetization $M_{x}\left( \hat{\rho}^{\prime}\right) \!=\! 0.5$ (those values are rounded with one decimal place in this example). Summarizing, the equations are
\begin{eqnarray*}
x_{2} &=&0\text{,} \\
x_{3} &=&0\text{,} \\
x_{1}-x_{4} &=&1\text{,} \\
x_{1}+x_{4} &=&1\text{.}
\end{eqnarray*}%
Solving this system of coupled equations, we find the correspondent density matrix%
\begin{equation}
\hat{\rho}_{0}=\left[ 
\begin{array}{cc}
1 & 0 \\ 
0 & 0%
\end{array}%
\right] \text{,} \label{EstadoInicialI1i2}
\end{equation}%
with its experimental counterpart being shown as a bar chart in Fig. \ref{fig:Rho21-00}, in which the left(right) bar chart corresponds to the real(imaginary) contribution of $\hat{\rho}_{0}$. Therefore we use similar procedure into the reconstruction of the density matrices of   other quantum states.
\begin{figure}[t!]
 \centering
  \includegraphics[scale=0.31]{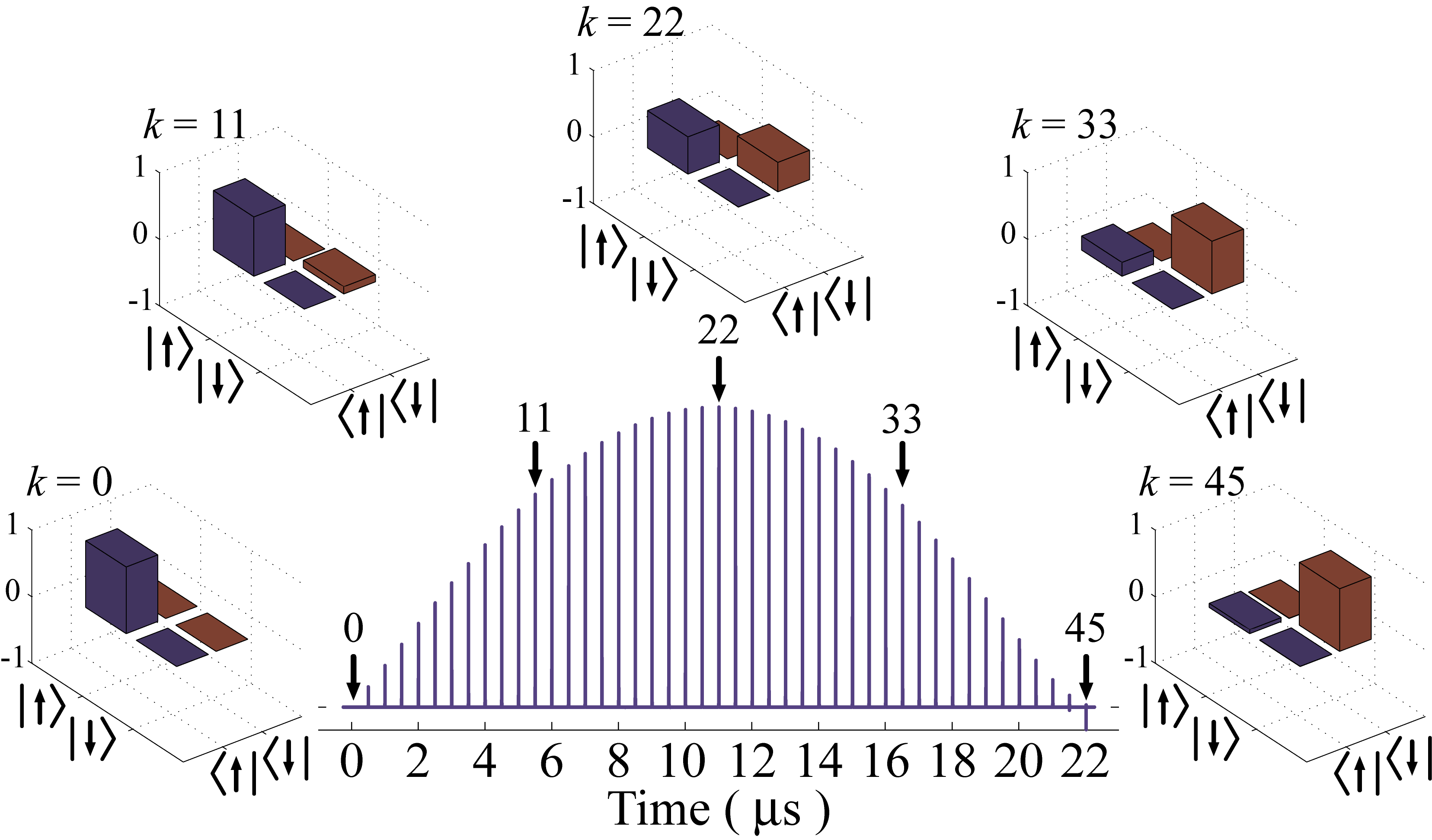}
\caption{(Color online) Spin dynamics under   $R_{-x}\left( \tau_{\textbf{r}}\right)$ rotation. The real part of five density matrices are displayed in the basis $\left\lbrace \left\vert \uparrow \right\rangle , \left\vert \downarrow \right\rangle\right\rbrace $
corresponding to the subscript $k=0,11,22,33,$ and $45$. In the center, the
NMR spectra correspond to the time dynamics of the magnetization $
M_{x}\left( \hat{\protect\rho }\right)$, in which the spectra marked with indexes $k=0,11,22,33,$ and $45$ are highlighted.} \label{fig:MatrizEspectros}
\end{figure}

\subsection{Implementation of the protocol to register the quantum dynamics}
The development of the protocol can be summarized as an initialization of the quantum state and the register of its quantum dynamics, those stages are depicted in Fig. \ref{fig:PulseSequence} as a pulse sequence.
 
\textit{Initialization :} The target quantum state obeys the definition of Eq. (\ref{pseudopurestate}) and follows the procedure similar to that used to obtain the quantum state of Eq.  (\ref{EstadoInicialI1i2}) at $t=0$.

\textit{Quantum dynamics :} The target quantum state is transformed using an operator defined by%
\begin{equation*}
R_{-x}\left( \tau _{%
\mathbf{r}}\right) =\exp \left[ i\omega _{1}\tau _{%
\mathbf{r}}\hat{\mathbf{I}}_{x}\right] \text{,}
\end{equation*}%
for different forty six values of $\tau _{\mathbf{r},k}$ $\in \left[ 0\ \mu 
\text{\texttt{s}},22\ \mu \text{\texttt{s}}\right] $ with $\bigtriangleup
\tau _{\mathbf{r}}=\tau _{\mathbf{r},k}-\tau _{\mathbf{r},k-1}=0.5\mu $%
\texttt{s}. In Fig. \ref{fig:MatrizEspectros} we show the dynamics of the nuclear spin encoded in forty six spectra (in the center of the figure) and only the real part of five density matrices represented by  bar charts. 

\section{Spin system $I=3/2$}
The NMR experiments are performed on a VARIAN INOVA 400 MHz spectrometer at the Institute of Physics in S\~{a}o Carlos (IFSC-USP). In this second setup we use Sodium dodecil sulfate sample (see Fig. \ref{fig:NMRscheme}(b)) in order to excite and to detect $^{23}$Na nuclei signals. The stoichiometry of this sample obeys $21.3$ wt \% of sodium dodecil sulfate, $3.7$ wt \% of decanol, and $75$ wt \% of deuterated water in a regime of a lyotropic liquid crystal. At strength static magnetic field of 9.39 Teslas, the Larmor frequency and quadrupolar couplings are $2\pi (105.842$MHz) and $2\pi (15$kHz), respectively. Also, $\pi $-pulse lengths of 8 $\mu $s and recycle delays of 500 ms are calibrated. The T$_{2}$ and T$_{1}$ relaxation times of $^{23}$Na nuclei are $2.6\pm 0.3$ ms and $12.2\pm 0.2$ ms, respectively.

At thermal equilibrium, most NMR systems are represented by their almost maximum mixture states, such as Eq.(\ref{densitymatrix}). Using suitable pulse sequences -- spin rotations and free evolutions --  any pseudo pure state is prepared \cite{Gershenfeld_1997,Cory_1997,Fortunato_2002} as depicted in Fig. \ref{fig:PulseSequence}. In this sense, the first dashed square of Fig.  \ref{fig:PulseSequence} represents the initialization procedure performed by strongly modulated pulses\cite{Fortunato_2002}, such that in this study we prepare the pseudo-nuclear spin coherent state \cite{Auccaise_2013} $\left\vert \psi(0) \right\rangle = \left\vert 3/2,+3/2\right\rangle
\equiv \left\vert \zeta \left( 0,0\right) \right\rangle $; the second square represents a hard pulse with variable length $\tau_{\textbf{r}}$ to control the spin rotation; the third dashed square represents the tomography pulse \cite{Teles_2007}. In particular, the initial quantum state means, from the NMR point of view, the precession of the magnetic moment around an axis defined by the orientation of the strong static magnetic field $B_{0}$. For more details on the experimental setup, the initialization of the pseudo-nuclear spin coherent state, and the register of the dynamics to flip the spin 3/2, see Ref. \cite{Auccaise_2013}. 


\end{document}